\pdfoutput=1 % For arXiv's AutoTeX
\documentclass[11pt]{article}

\usepackage[utf8]{inputenc} % UTF-8 character support

\usepackage{amsmath}
\usepackage{amsthm}         % theorem and proof environments
\usepackage{bm}             % Bold type for vectors
\usepackage[style=authoryear-ibid]{biblatex}
\usepackage{csquotes}       % uniform quote style with \enquote, cited quotes with \textcquote
\usepackage[T1]{fontenc}    % Support more of Palatino's glyphs
\usepackage{import}         % https://www.overleaf.com/learn/latex/Management_in_a_large_project#Using_the_import_package
\usepackage{mathtools}      % \cramped, \mathclap, \begin{multlined}, \coloneqq, et al
\usepackage{microtype}      % improved spacing and justification of paragraphs
\usepackage{mleftright}     % \mleft, \mright improve spacing around \left, \right
\usepackage[                % Palatino text fonts
	largesc,                  % small caps is 8% larger than default petite caps
	osf,                      % old-style (aka lower case) figures (numerals 0-9)
	proportional,             % proportional rather than tabular alignment for figures (numerals 0-9)
	theoremfont,              % Upright numbers and parentheses amidst italic theorem text
	trueslanted,              % \textsl is true slanted rather than italic with upright punctuation
]{newpxtext}
\usepackage[                % Matching Palatino math fonts. Must be loaded after newpxtext.
	vvarbb,                   % Use STIX font's blockboard bold, recommended by newpx documentation
]{newpxmath}
\usepackage{newunicodechar} % \newunicodechar for Unicode input (eg `≤` for `\le`)
\usepackage{relsize}        % \textlarger/\textsmaller
\usepackage{xcolor}         % For colored hyperlink definitions in \hypersetup below

\usepackage[                % Adobe Source Code Pro typewriter font
	scale=.96,                % Scaled to 96% size to match Palatino's ex height
]{sourcecodepro}            % Must be loaded after newpxtext

\AtBeginDocument{\edef\theoremfontfamily{\thdefault}}
\makeatletter
\DeclareRobustCommand*{\abbr}[1]{%
	\texorpdfstring{%
		{%
			\let\&\smallamp%
			\ifx\f@family\theoremfontfamily%
				\normalfont\itshape%
				\ifx\f@series\bfdefault%
					\bfseries%
				\fi%
			\fi%
			\scshape%
			\selectfont \MakeLowercase{#1}%
		}%
	}{#1}% Can't use \MakeUppercase here as we'd like. https://tex.stackexchange.com/a/417078
}
\makeatother
\newcommand*\smallamp{\textsmaller[2]{\char"26}}

\renewcommand{\vec}{\bm}             % Bold Roman type for vectors
\newcommand{\R}{\mathbb{R}}          % Real numbers
\newcommand{\Graphs}[2][]{\mathcal{G}_{#2%
	\if\relax\detokenize{#1}\relax%      https://tex.stackexchange.com/a/53091
	\else%
		\unspace ,#1%
	\fi%
}}
\newcommand{\N}{\mathbb{N}}          % Natural numbers
\let\Pr\undefined                                % Undefine default \Pr so we can redfine it
\DeclareMathOperator{\Pr}{\mathbb{P}}            % Re-define it the same way amsmath does, but with \mathbb
\DeclareMathOperator{\E}{\mathbb{E}} % Expectation operator
\DeclareMathOperator{\indicate}{\mathbb{1}}% Indicator function
\newcommand{\BigO}{\mathcal{O}}      % Big-Oh notation
\newcommand{\Dyads}[1]{\mathcal{D}_{#1}} % Set of dyads on vertex set [#1]
\newcommand{\given}{\ifnum\currentgrouptype=16 \mathrel{}\middle|\mathrel{}\else\mid\fi}

\DeclareFieldFormat{eprint:ssrn}{%
  \mkbibacro{SSRN}\addcolon\space
  \ifhyperref
    {\href{https://ssrn.com/abstract=#1}{\nolinkurl{#1}}}
    {\nolinkurl{#1}}}
\DefineBibliographyExtras{american}{
	\DeclareBibstringSet{latin}{
		andmore, andothers, confer, ibidem,
		idem, idempf, idempm, idempn, idempp, idemsf, idemsm, idemsn,
		loccit, opcit, passim, sequens, sequentes,
	}%
	\DeclareBibstringSetFormat{latin}{\mkbibemph{#1}}%
}
\UndefineBibliographyExtras{american}{
	\UndeclareBibstringSet{latin}
}
\NewBibliographyString{
	appendix,appendixs, % Pagination for pin cites to specific appendices
	chapters, % `chapter` already defined as "chap." in biblatex's american.lbx
	commenton, % For the `relatedtype` field. Also available from the `apa` style
	presentation, % For the `type` field
	regulation,regulations, % Pagination for UK statutory instruments
	slide,slides, % Pagination for `presentation` type entries
}
\DefineBibliographyStrings{american}{%
	appendix    = {app\adddot},
	appendixs   = {apps\addot},
	chapters    = {chaps\adddot},
	commenton   = {comment\space on},
	presentation= {presentation},
	regulation  = {reg\adddot},
	regulations = {regs\adddot},
	slide       = {slide},
	slides      = {slides},
}

\SetCiteCommand{\autocite}
\SetBlockEnvironment{quotation}

\mleftright

\newunicodechar{π}{\pi}
\newunicodechar{∼}{\sim}
\newunicodechar{μ}{\mu}
\newunicodechar{≥}{\ge}
\newunicodechar{≠}{\ne}

\newcommand{\St}{\mathscr{S}}        % State space (arbitrary set)
\newcommand{\perm}{\pi}              % Arbitrary (set of) permutation(s) on \St
\newcommand{\lnpart}{\zeta}          % log partition function
\newcommand{\degseq}{\vec{\beta}}    % degree sequence of a graph
\newcommand{\slfrac}[2]{\left.#1\middle/#2\right.}
\newcommand{\Z}{\mathbb{Z}}          % Integers

\usepackage[
	letterpaper,
	margin = 1in,
]{geometry}
\usepackage[inline]{enumitem}

\usepackage{multicol}       % For multicolumn lists

\usepackage[nospace]{varioref}      % \vref for "on the preceeding page" references
\usepackage[pdfusetitle]{hyperref}  % Hyperlinks
\usepackage[noabbrev]{cleveref}     % Context-aware internal cross referencing; noabbrev for SJS style

\Crefname{subsection}{Subsection}{Subsections}
\Crefname{subsubsection}{Sub-subsection}{Sub-subsections}
\theoremstyle{plain}
\newtheorem{theorem}{Theorem}[section]  % Restart counter in each section
\newtheorem{proposition}[theorem]{Proposition}
\newtheorem{lemma}[theorem]{Lemma}
\newtheorem{corollary}[theorem]{Corollary}
\theoremstyle{definition}
\newtheorem{definition}[theorem]{Definition}
\newtheorem{example}[theorem]{Example}

\newtheorem*{examplecontinued}{Example \continuation}
\newcommand\continuation{\usecontinuationonlyviaExampleContinuedenvironment}
\newenvironment{ExampleContinued}[1]
	{\renewcommand{\continuation}{\ref{#1}}\begin{examplecontinued}[continued]}
	{\end{examplecontinued}}

\theoremstyle{remark}
\newtheorem{observation}[theorem]{Observation}
\crefname{observation}{observation}{observations}
\Crefname{observation}{Observation}{Observations}

\makeatletter
\newcommand*\introduce[1]{%
	\ifx\f@family\theoremfontfamily%
		{\normalfont\bfseries\selectfont #1}%
	\else%
		\textbf{\emph{#1}}%
	\fi%
}
\makeatother

\newlist{conditions}{enumerate}{1}
\setlist[conditions,1]{
	label      = (\alph*),
	before     = {\raggedcolumns\begin{multicols}{2}},
	after      = {\end{multicols}\flushcolumns},
	widest*    = 1, % Make the width of the label just big enough for the first label ('a' in this case)
	leftmargin = 0pt,%Flush 2nd and subsequent lines of each item to left edge of surrounding text
	itemindent = *, % Flush label to left edge of surrounding text
}
\crefname{conditionsi}{condition}{conditions}
\Crefname{conditionsi}{Condition}{Conditions}

\newcommand\email[1]{\href{mailto:#1}{\nolinkurl{#1}}}
\hypersetup{
	bookmarksopen=true,
	bookmarksnumbered=true,
	colorlinks,
	linkcolor={red!50!black},  % 50% black, 50% red
	citecolor={blue!50!black}, % 50% black, 50% blue
	urlcolor={blue!80!black},  % 20% black, 80% blue
}

\usepackage{appendix}       % To show an "Appendices" heading before appendices
\ifdefined\TitlePage
	\usepackage{multirow}       % \multirow in tables
	\usepackage[figure,table]{totalcount} % count number of figures, tables
\fi

\setlist[enumerate,1]{label=(\alph*)}
\setlist[enumerate,2]{label=(\roman*)}

\title{Longitudinal Network Models and Permutation-Uniform Markov Chains%
	\texorpdfstring{%
		\footnote{%
			This article—which is the accepted version of the following published article—may be used for non-commercial purposes in accordance with the \href{http://www.wileyauthors.com/self-archiving}{Wiley Self-Archiving Policy}.
			\begin{displayquote}[][.]\fullcite{self}\end{displayquote}%
		}%
	}{}%
}
\newcommand*{\RunningTitle}{Markov Chains of Networks}
\newcommand*{\affil}[3][]{#2 \\ \emph{\textsmaller{%
	\if\relax\detokenize{#1}\relax%      https://tex.stackexchange.com/a/53091
	\else%
	\fi%
	\space#3%
}}}
\date{December 2022}
\author{%
	\texorpdfstring{
		\affil{William K. Schwartz}{Secretariat Economists \abbr{LLC}}\and%
		\affil[Department of Applied Mathematics]{Sonja Petrovi\'c}{Illinois Institute of Technology}\and%
		\affil[Department of Applied Mathematics]{Hemanshu Kaul}{Illinois Institute of Technology}
	}{William K. Schwartz, Sonja Petrović, and Hemanshu Kaul}%
}
\begin{document}

% Title page
\ifdefined\TitlePage
	%TC:ignore
	\pagenumbering{roman}
	\pdfbookmark{Title Page}{sec:title-page}
	\import{modules/article}{\TitlePage}
	\pagenumbering{arabic}
	%TC:endignore
\fi

\pdfbookmark{\RunningTitle}{article-top}
\maketitle

\begin{abstract}
	Consider longitudinal networks whose edges turn on and off according to a discrete-time Markov chain with exponential-family transition probabilities.
	We characterize when their joint distributions are also exponential families with the same parameter, improving data reduction.
	Further we show that the permutation-uniform subclass of these chains permit interpretation as an independent, identically distributed sequence on the same state space.
	We then apply these ideas to temporal exponential random graph models, for which permutation uniformity is well suited, and discuss mean-parameter convergence, dyadic independence, and exchangeability.
	Our framework facilitates our introducing a new network model; simplifies analysis of some network and autoregressive models from the literature, including by permitting closed-form expressions for maximum likelihood estimates for some models; and facilitates applying standard tools to longitudinal-network Markov chains from either asymptotics or single-observation exponential random graph models.

	\medskip\noindent
	\begin{itemize*}[
		before   = {\textbf{Keywords:}},
		label    = {},
		itemjoin = {; },
		after    = {.},
	]
		\item conditional exponential families
		\item compression
		\item data reduction
		\item dyadic independence
		\item \abbr{ERGM}s
		\item exponential families
		\item longitudinal networks
		\item Markov chains
		\item permutation uniformity
		\item temporal exponential random graph models
	\end{itemize*}
\end{abstract}
%\tableofcontents

\section{Introduction}\label{sec:intro}
	Over the last half century, the statistical modeling of networks has ramified.
	One branch of studies has included \emph{exponential random graph models} (\abbr{ERGM}s), which have proved successful for modeling single observations of random networks.
	Their probabilities are \emph{exponential families}, parameterized functions of certain observables, called \emph{sufficient statistics}, such as the number of edges, the degree sequence, or the counts of specified subgraphs.
	Another branch has modeled \introduce{longitudinal} or \introduce{dynamic networks}, in which edges blossom and die but nodes stay the same \autocites(\bibstring{see}, e.g.,)(){Holland:1977fi}{frank91}{Snijders2005}, as discrete-time Markov chains \autocite[\autocap{f}irst discussed in][]{katz-proctor-1959} where each row of the matrix of transition probabilities has an exponential-family representation with the same parameter, as in \textcite{Robins:2001ke}.
	\Textcite{hfx2010} explored this family of models, calling them \introduce{temporal \abbr{ERGM}s} (\abbr{TERGM}s), and focused on the dyadic independence case.

	The present study asks, \emph{When are these types of discrete-time Markov chains of networks exponential families with the same parameter as their transition matrices?}
	Our motivation is to re-cast \abbr{TERGM}s as \abbr{ERGM}s since the latter are generative statistical models, allowing scientists to estimate parameters that weight the importance of different sufficient statistics in explaining why a network has a certain topology.
	Moreover, \cref{sec:cef} discusses the extended scope for data compression that arises when a \abbr{TERGM}'s joint distribution has an exponential family representation.
	Finally, statisticians have developed sophisticated techniques for estimating these parameters and testing goodness of fit for \abbr{ERGM}s \autocites(\bibstring{see}, e.g.,)(){goldenberg2010}{kolaczyk-2017}[88--97]{kolaczyk-2020-ch6}.

	The answer is, perhaps surprisingly, not \enquote{always}.
	Our \vref{thm:mef-exp-fam} implies that \abbr{TERGM}s have an exponential family joint distribution with the same parameter as their transition matrices if and only if the normalizing term in their transition matrices is the same across all rows.
	A sufficient condition for this tractable class of \abbr{TERGM}s arises in the case that the Markov chain is also \emph{permutation-uniform} (\emph{p-uniform}), meaning that every row of the transition matrix is a permutation of every other row.

	The key insight is that when a Markov chain is p-uniform, composing those permutations with the Markov chain itself produces an independent and identically distributed (\abbr{IID}) sequence on the same state space.
	Readers familiar with \textcite{diaconis-freedman-1999} may recognize p-uniform Markov chains as being induced by sets of permutations on the state space itself.
	When coupled with existing exponential family models of networks, the \abbr{IID} sequence's finite-sample joint distribution maintains the exponential family representation with a low-dimensional sufficient statistic interpretable in terms of the underlying Markov chain.
	This is similar to the autoregressive model of networks in \textcite{tensor-regression-panel-data}; see \cref{ex:modular-autoregression,ex:var}.
	Moreover, the \abbr{IID} sequence can be viewed as a single observation of a multigraph drawn from an \abbr{ERGM} whose parameter estimates and goodness-of-fit tests apply to the original Markov chain.

	In \cref{sec:puniformity}, the main \lcnamecref{thm:uniform-chain}, \cref{thm:uniform-chain}, establishes the key identification of p-uniform Markov chains with an \abbr{IID} sequence on the same state space.
	This identification plays nicely with exponential family representations of the transition matrix and the interpretability of its sufficient statistic.
	In this way, we can translate certain Markov chains to \abbr{IID} sequences, perform statistical analysis on the latter, and draw conclusions about the former.
	Applications of p-uniform Markov chains include autoregressive processes on discrete state spaces and several of the examples in \cref{sec:examples-lit}.
	\Textcite[\S~6]{consistency-exp-fam-mle} called for ways of analyzing independent random variables in place of dependent random variables in Markov chains of networks.
	The novelty of the techniques in this paper are disposing of the temporal dependence in a Markov chain, and maintaining interpretability of parameters and sufficient statistics while doing so.
	For certain Markov chains (see \cref{sec:examples-lit}), dispensing with temporal dependence allows us to compute maximum-likelihood estimators (\abbr{MLE}s) from closed-form expressions where previously the literature contemplated only Markov-chain Monte Carlo (\abbr{MCMC}) or Newton's method algorithms.

	In \cref{sec:interp}, we apply the theories of exponential families of transition matrices and of p-uniform Markov chains to network models in the context of the statistical independence of the random connections in the networks.
	The main result in \cref{thm:suff-stat-dyadic-indep-multigraphs,thm:suff-stat-dyadic-indep-multigraph-joint-prob} is that we may replace $t$ observations of certain p-uniform Markov chains of graphs with a single observation of a corresponding multigraph.
	We introduce \emph{exponential random $t$-multigraph model}s ($t$-\abbr{ERMGM}s) for this purpose.
	We expect much of existing \abbr{ERGM} theory to apply without significant modification to $t$-\abbr{ERMGM}s: \textcite{mle-log-linear-models} thoroughly described parameter estimation, and \textcite{Gross2022} surveyed the literature for goodness-of-fit testing.
	\Cref{sec:examples-lit,sec:loyalty-tergm} illustrate our results with existing Markov-chain network models as well as a new model.
	The former \lcnamecref{sec:examples-lit} examines models from \textcite{hfx2010}: their density and stability models are p-uniform, but their reciprocity and transitivity models are not.
	The latter \lcnamecref{sec:loyalty-tergm} offers a novel network model whose parameters, one per actor in the network, may be interpreted to measure loyalty in terms of the tendency of each actor to change ties from one time to the next.

	Finally, \cref{sec:conclusion} concludes with a summary of how all our results fit together to constitute an analytical framework for certain kinds of \abbr{TERGM}s.

	After this introduction and some notation in \cref{sec:notation}, \cref{sec:cef} identifies which Markov chains have a joint distribution with a low-dimensional sufficient statistic.
	This is particularly important when the state space is as large as the set of all $\BigO(\cramped{2^{n^2}})$ networks on $n$ vertices.
	The answer lies in exponential families: \cref{thm:mef-exp-fam} is a Darmois-Koopman-Pitman-type theorem for discrete-time Markov chains on discrete state spaces.

	\subsection{Notation and Terminology}\label{sec:notation}
		First let us dispense with some bookkeeping notation.
		Denote the set $\{0, 1, 2, \dotsc\}$ of natural numbers by $\N$ and its positive subset by $\N_{>0}$.
		$\vec{1}$ is the vector of all ones of whatever dimension is appropriate.
		If $n \in \N$, then $[n] \coloneqq \{1, \dotsc, n\}$.

		Now we turn to statistics terminology.
		Let $\St$ be a discrete (i.e., at most countable) state space.
		Where needed, we endow discrete spaces with the discrete (power set) $\sigma$-algebra, allowing us to brush aside abstruse considerations of measurability.
		Let $X \coloneqq \{X_t\}_{t\in\N}$ be an $\St$-valued stochastic process, and let $\Theta \subseteq \R^d$, for some $d \ge 1$, be a non-empty parameter space.
		We assume the existence of a set of probability measures $\Pr_{\vec{\theta}}$, for each $\vec{\theta} \in \Theta$, under each of which $X$ is a time-homogeneous Markov chain with transition matrix $P_{\vec{\theta}} \in \R^{\St \times \St}$.
		Denote $\mathscr{P} \coloneqq \{P_{\vec{\theta}}\}_{\vec{\theta} \in \Theta}$.
		When we say \emph{Markov chain}, we always assume temporal homogeneity.
		We typically write an entry of $P_{\vec{\theta}}$ as $P_{\vec{\theta}}(a, b)$ for $a, b \in \St$.
		For all $\vec{\theta} \in \Theta$, $t \in \N_{>0}$, and $x_0 \in \St$, define $\Pr^t_{\vec{\theta}}( \cdot \given X_0 = x_0)$ to be the \introduce{restriction} of $\Pr_{\vec{\theta}}( \cdot \given X_0 = x_0)$ to the $\sigma$-algebra generated by $X_0, \dotsc, X_t$.
		Further, define $L^t_{\vec{\theta},x_0}$ to be the probability mass function (\abbr{PMF}) of the law of $X_1, \dotsc, X_t$ under $\Pr^t_{\vec{\theta}}( \cdot \given X_0 = x_0)$.

		A parameterized set $\mathscr{M} \coloneqq \{\mu_{\vec{\theta}}\}_{\vec{\theta} \in \Theta}$ of \abbr{PMF}s on $\St$ is an \introduce{exponential family} if there exist $\ell \in \N_{>0}$ and functions $\lnpart \colon \Theta \to \R$, $\kappa \colon \St \to [0, \infty)$, $\vec{\eta} \colon \Theta \to \R^\ell$, and $\vec{\tau} \colon \St \to \R^\ell$ such that, for all parameters $\vec{\theta} \in \Theta$, the probability mass of any state $a \in \St$ under $\mu_{\vec{\theta}}$ is
		\begin{equation}\label{eq:exp-fam}
			\mu_{\vec{\theta}}(a)
			= \kappa(a)\exp\left(\vec{\eta}(\vec{\theta}) \cdot \vec{\tau}(a) - \lnpart(\vec{\theta})\right).
		\end{equation}
		\Cref{eq:exp-fam} is an \introduce{exponential family representation} of $\mathscr{M}$, $\lnpart$ the \introduce{log-partition function} and thus $e^{\lnpart}$ the \introduce{partition function} \autocite[Eq.~3.6]{wainwright-jordan}, $\kappa$ the \introduce{carrier measure}, $\eta$ the \introduce{parameter function}, and $\vec{\tau}$ a sufficient statistic for $\vec{\theta}$.

		$\mathscr{M}$'s \introduce{natural parameter space} is $\{\vec{\gamma} \in \R^\ell \given \sum_{b\in\St}\kappa(b)e^{\vec{\gamma} \cdot \vec{\tau}(b)} < \infty\}$.
		We always stipulate that $\vec{\eta}(\Theta)$ lies in the natural parameter space \autocites[114]{casella-stat-infer}; hence $|\lnpart(\vec{\theta})| < \infty$ for all $\vec{\theta} \in \Theta$.

		The parameterized set $\mathscr{P} = \{P_{\vec{\theta}}\}_{\vec{\theta} \in \Theta}$ of transition matrices is a \introduce{conditional exponential family} (\abbr{CEF}) if each row $\{P_{\vec{\theta}}(a, \cdot)\}_{\vec{\theta} \in \Theta}$, where $a \in \St$, has an exponential family representation with the same parameter function $\vec{\eta} \colon \Theta \to \R^\ell$ \autocite[Def.~A]{conditional-exponential-families}.
		Specifically, $\mathscr{P}$ is a \abbr{CEF} if there exist functions $\lnpart \colon \St \times \Theta \to \R$, $\kappa \colon \St \times \St \to [0, \infty)$, $\vec{\tau}\colon \St \times \St \to \R^\ell$ such that
		\begin{equation}\label{eq:def-cef}
			P_{\vec{\theta}}(a, b)
			= \kappa(a, b)\exp\left(\vec{\eta}(\vec{\theta}) \cdot \vec{\tau}(a, b) - \lnpart(a, \vec{\theta})\right)
		\end{equation}
		for all $a, b \in \St$ and all $\vec{\theta} \in \Theta$.
		We call \cref{eq:def-cef} a \introduce{\abbr{CEF} representation} of $\mathscr{P}$.
		Temporal \abbr{ERGM}s are Markov chains on the set of a networks with a given number of nodes whose transition probabilities have a \abbr{CEF} representation.

		$\mathscr{P}$'s \introduce{natural parameter space} is $\{\vec{\gamma} \in \R^\ell \given \sum_{b \in \St} \kappa(a, b)e^{\vec{\gamma} \cdot \vec{\tau}(a, b)} < \infty \text{ for all } a \in \St\}$.
		We always stipulate that $\vec{\eta}(\Theta)$ lies in the natural parameter space \autocite[598]{conditional-exponential-families}; hence $|\lnpart(a, \vec{\theta})| < \infty$ for all $\vec{\theta} \in \Theta$ and all $a \in \St$.

		The \abbr{CEF} is a \introduce{conditionally additive exponential family} (\abbr{CAEF}) if $\lnpart(a, \vec{\theta}) = \psi(a)\phi(\vec{\theta})$ for some functions $\phi \colon \Theta \to \R$ and $\psi \colon \St \to \R$ \autocite[Def.~B]{conditional-exponential-families}.
		Finally, the \abbr{CAEF} is a \introduce{Markovian exponential family} (\abbr{MEF}) if $\psi(a)$ is a non-zero constant for all $a \in \St$, i.e., $\lnpart(a, \vec{\theta}) = \lnpart(b, \vec{\theta}) \eqqcolon \lnpart(\vec{\theta})$ so that \autocite[Eq.~2.1]{Hudson:1982hb}, for all $a, b \in \St$ and all $\vec{\theta} \in \Theta$,
		\begin{equation}\label{eq:def-mef}
			P_{\vec{\theta}}(a, b)
			= \kappa(a, b)\exp\left(\vec{\eta}(\vec{\theta}) \cdot \vec{\tau}(a,b) - \lnpart(\vec{\theta})\right).
		\end{equation}
		We call \cref{eq:def-mef} an \introduce{\abbr{MEF} representation} of $\mathscr{P}$.
		\Cref{eq:def-cef,eq:def-mef} defining \abbr{CEF}s and \abbr{MEF}s differ only in that the log-partition function for the \abbr{CEF} in \cref{eq:def-cef} depends on which row of the transition matrix it is in, whereas the log-partition function for the \abbr{MEF} in \cref{eq:def-mef} is the same in every row of the transition matrix.

		\begin{example}[{\citereset\Cite[Example \RN{4}.3$\cdot$3, \pnfmt{358--359}]{Gani:1955gp}}]\label{ex:scalar-param-mef}
			\pushQED{\qed}
			The following is an example of an \abbr{MEF} with a scalar parameter.
			Set $\St \coloneqq \{1, 2, 3\}$, $\Theta \coloneqq (0, \infty)$, $\eta(\theta) \coloneqq \log(\theta)$,
			\begin{align*}
				\tau &\coloneqq
				\begin{bmatrix}
					1 & 1 & 3 \\
					3 & 3 & 1 \\
					1 & 3 & 3
				\end{bmatrix}, &\text{and}
				&&
				\kappa &\coloneqq
				\begin{bmatrix}
					2            & 1           & 1 \\
					\frac{1}{3}  & \frac{2}{3} & 3 \\
					\frac{11}{4} & 1           & \frac{1}{4}
				\end{bmatrix}, &\text{so}
				&&
				P_{\theta} &=
				\frac{1}{3\theta + \theta^3}
				\begin{bmatrix}
					2           \theta   &            \theta   &            \theta^3 \\
					\frac{1}{3} \theta^3 & \frac{2}{3}\theta^3 & 3          \theta  \\
					\frac{11}{4}\theta   &            \theta^3 & \frac{1}{4}\theta^3
				\end{bmatrix}.
				\qedhere
			\end{align*}
			\popQED
		\end{example}

		\Textcite{heyde-feigin-1975} first introduced \abbr{CAEF}s, which \textcite{conditional-exponential-families} generalized to \abbr{CEF}s.
		\Textcite{Hudson:1982hb} coined the name for \abbr{MEF}s.
		None were aware of the analysis in \textcite{Gani:1955gp} decades earlier of what turned out to be \abbr{MEF}s.

	\subsection{Markov Chains with Low-Dimension Sufficient Statistics}\label{sec:cef}
		\introduce{Darmois-Koopman-Pitman} (\abbr{DKP}) theorems assert the uniqueness of exponential families as the only parameterized families of probability distributions whose sufficient statistics do not grow in dimension with the size of the sample (under some conditions).
		Continuous state spaces have been the usual focus of textbook renditions, e.g., \textcite[Thm.~6.18, \pnfmt{40}]{lehmann-casella-point-est}.
		Discrete analogues in the research literature have included \textcites[\S~\RN{4}]{Gani:1955gp}{Andersen:1970bg,Denny:1972ha,diaconis-freedman-exch-1984}.
		The former gave three such results for scalar parameters, the Markov-chain version of which we discuss below because it motivates this \lcnamecref{sec:cef}'s main result, \cref{thm:mef-exp-fam}.

		That \lcnamecref{thm:mef-exp-fam} says roughly that a Markov chain's parameterized family of finite-sample joint \abbr{PMF}s have an exponential family representation if and only if the corresponding set of transition matrices have an \abbr{MEF} representation.
		As a consequence, $\mathscr{P}$ is a \abbr{CEF} that gives rise to an exponential family of joint \abbr{PMF}s if and only if $\mathscr{P}$ is an \abbr{MEF}.
		In that case, the joint distribution and transition matrix share the same parameter and the sufficient statistics $\vec{\tau}(X_i, X_{i+1})$ for the transitions sum to form the sufficient statistic $\sum_{i=0}^{t-1}\vec{\tau}(X_i, X_{i+1})$ for the joint distribution.
		The dimension of both statistics is $\ell$, which could be much smaller than the dimension of a sufficient statistic for the joint distribution's parameter in the general case.

		In the general case—even when $\mathscr{P}$ is not a \abbr{CEF}—Markov chains' joint distributions form exponential families, but with a high-dimensional sufficient statistic.
		A statistic of $(X_1, \dotsc, X_t)$ that is sufficient for the entries of $P_{\vec{\theta}}$ is the \introduce{transition-count matrix} $N_t \in \N^{\St \times \St}$ whose $a,b$ entry,
		\begin{equation*}
			N_t(a,b) \coloneqq \sum_{i=0}^{t-1}\indicate(X_i = a)\indicate(X_{i+1} = b),
		\end{equation*}
		where the random variable $\indicate(X_i = a)$ is one when $X_i = a$ and zero otherwise, counts the number of times $X$ transitioned from state $a$ to state $b$ by time $t$ \autocites(\bibstring{confer})()[Eqs.~8--9]{Gani:1955gp}[Eq.~7]{Stefanov:1995cl}.
		The $a$th entry of $N_t \vec{1}$ gives the number of times that $X$ has visited state $a$ during times $0, \dotsc, t-1$, inclusive.
		Even if $\St$ is finite, the number of degrees of freedom of the minimal representation of $N_t$ is $|\St|^2 - |\St|$ \autocite[\S~2]{Stefanov:1991cl}.
		That is $\BigO(\cramped{2^{n^2}})$ if $\St$ is the set of graphs on the vertex set $[n]$.
		$N_t$ is sufficient for the entries of $P_{\vec{\theta}}$ because the joint \abbr{PMF} of $(X_1, \dotsc, X_t)$ is
		\[
			L^t_{\vec{\theta}, X_0}\left(X_1, \dotsc, X_t\right)
			= \exp\left(\sum_{a, b\in \St} N_t(a,b)\log P_{\vec{\theta}}(a, b)\right)
		\]
		(as long as we take $0 \log 0 \coloneqq 0$) \autocite[Eq.~3.2.5]{Kuchler:1997vo}.

		Discrete-\abbr{DKP}-type results offer a way to fend off $N_t$'s dimensional onslaught.
		In other words, \abbr{MEF}s afford greater scope for data compression than \abbr{CEF}s.
		Among the earliest such results in the literature were \textcites[§~\RN{4}.3]{Gani:1955gp}{Gani:1956ip}.
		In the former, \citeauthor{Gani:1955gp} assumed that $\St$ is finite, $\Theta$ is a set of scalars in $\R$, and that every transition matrix $P_\theta \in \mathscr{P}$ has a stationary distribution \autocite[fn.~9 on \pno~15]{Schwartz2021}	and is differentiable with respect to $\theta$ \autocite[\autocap{t}hough differentiability is inessential---\bibstring{see}][fn.~8 on \pno~14]{Schwartz2021}.
		He concluded (in slightly different language) that $X$'s finite-sample joint \abbr{PMF} admits a one-dimensional statistic sufficient for $\theta$ if and only if $\mathscr{P}$ is an \abbr{MEF}.

		\Cref{thm:mef-exp-fam} drops \citeauthor{Gani:1955gp}'s ergodicity, scalar parameter, and other conditions in the sense that it says that \abbr{MEF}s and only \abbr{MEF}s make \emph{any} (discrete space and time homogeneous) Markov chain's finite-sample joint \abbr{PMF}s an exponential family.
		Hence the \lcnamecref{thm:mef-exp-fam} gives a sufficient condition—but no necessary conditions—for the chain's joint \abbr{PMF} to admit an $\ell$-dimensional statistic sufficient for $\vec{\theta}$.

		The \lcnamecref{thm:mef-exp-fam}, whose measure theoretic proof we omit, is a consequence of \textcite[Cor.~6.3.4]{Kuchler:1997vo}.
		The analogy between our notation and theirs \autocite[19,65--68]{Kuchler:1997vo} is easiest to see by reading our $\Pr_{\vec{\theta}}( \cdot \given X_0 = x_0)$ as their probability measure $Q_{\vec{\theta}, x_0}$, for which $Q_{\vec{\theta}, x_0}(X_0 = x_0) = 1$ and $P_{\vec{\theta}}(x_0, x_1) = Q_{\vec{\theta}, x_0}(X_1 = x_1)$.

		\begin{theorem}\label{thm:mef-exp-fam}
			$\mathscr{P}$ is the \abbr{MEF} from \cref{eq:def-mef} with $\kappa(a, b) = P_{\vec{\theta}_0}(a, b)$ for some $\vec{\theta}_0 \in \Theta$ and all $a, b \in \St$ if and only if $\kappa(a, b) = L^1_{\vec{\theta}_0, a}(b)$ for some $\vec{\theta}_0 \in \Theta$ and all $a, b \in \St$, and $\{L^t_{\vec{\theta},x_0}\}_{\vec{\theta} \in \Theta}$ is the exponential family on $\St^t$ such that
			\[
				L^t_{\vec{\theta},x_0}(x_1, \dotsc, x_t)
				= \exp\left(
					\vec{\eta}(\vec{\theta}) \cdot \sum_{i=0}^{t-1}\vec{\tau}(x_i, x_{i+1}) - t\lnpart(\vec{\theta})
				\right)\prod_{i=0}^{t-1}\kappa(x_i, x_{i+1})
			\]
			for all $x_0, x_1, \dotsc, x_t \in \St$, $t \in \N_{>0}$, and $\vec{\theta} \in \Theta$.
		\end{theorem}

		While \cref{thm:mef-exp-fam} gives a general way to recognize when families of transition matrices \emph{are} \abbr{MEF}s, the next \lcnamecref{thm:mef-uniqueness-of-suff-stats}, due to \citeauthor{Gani:1955gp}, helps us in \cref{sec:examples-lit} to recognize when (scalar parameterized) \abbr{CEF}s \emph{are not} \abbr{MEF}s.
		It says the rows of a scalar sufficient statistic in an \abbr{MEF} all contain the same set of numbers.
		For \vref{ex:scalar-param-mef}, that set is $\{1, 3\}$.
		We include a sketch of \citeauthor{Gani:1955gp}'s proof because the main idea is simple and instructive \autocites(\autocap{s}imilar uses of the idea appeared in)()[454]{bhat-gani-1960}[\autocap{a}nd in][Example 2]{Bhat1988a}.
		The \lcnamecref{thm:mef-uniqueness-of-suff-stats} foreshadows \cref{sec:puniformity} in which we study a stronger condition that requires every row of a transition matrix to be a permutation of every other row.

		\begin{proposition}[{\citereset\Cite[357--358]{Gani:1955gp}}]\label{thm:mef-uniqueness-of-suff-stats}
			For the \abbr{MEF} in \cref{eq:def-mef}, suppose $\ell = 1$ (i.e., $\eta(\vec{\theta})$ is a scalar), $\eta(\Theta)$ contains a number other than zero, and $a, c \in \St$.
			Then $\{\tau(a, b) \given b \in \St\} = \{\tau(c, b) \given b \in \St\}$.
		\end{proposition}

		\begin{proof}[Proof sketch.]
			Since $\sum_{b \in \St} P_{\vec{\theta}}(a, b) = 1$, we have, for all $\vec{\theta} \in \Theta$,
			\[
				\exp(\lnpart(\vec{\theta}))
				= \sum_{b \in \St} \kappa(a, b)\exp(\eta(\vec{\theta})\tau(a, b))
				= \sum_{b \in \St} \kappa(c, b)\exp(\eta(\vec{\theta})\tau(c, b)).
			\]
			Exponential functions $x \mapsto e^{r x}$ are linearly independent for different scalars $r$ \autocite{axler-linear-independence-of-exp-funcs}, so we can set some of the coefficients and exponents equal when $\eta(\vec{\theta}) \ne 0$.
		\end{proof}

		We conclude this \lcnamecref{sec:cef} with the following statistical \lcnamecref{thm:mef-expected-suff-stat} for \abbr{MEF}s.
		It disposes of temporal dependence when we just need the \introduce{mean parameter} $\E[\vec{\tau}(X_t, X_{t+1})]$, which is crucial for \abbr{MLE} \autocites{mle-log-linear-models}[Thm.~6.2, \pnfmt{125--126}]{lehmann-casella-point-est}.

		\begin{theorem}\label{thm:mef-expected-suff-stat}
			Let $\mathscr{P}$ be the \abbr{MEF} from \cref{eq:def-mef}.
			Let $\vec{\theta} \in \Theta$ such that $\vec{\eta}(\vec{\theta})$ is in the interior of the natural parameter space and $\vec{\eta}$ is differentiable at $\vec{\theta}$ with Jacobian matrix $J_{\vec{\theta}}$.
			Then, for any $t \in \N$, $\E_{\vec{\theta}}[\vec{\tau}(X_t, X_{t+1})] = J^{-1}_{\vec{\theta}} \nabla \lnpart(\vec{\theta})$.
		\end{theorem}

		A counterpart to this result is \vref{thm:puniform-as-convergence}.

		\begin{proof}
			By \textcite[Thm.~1(i)]{conditional-exponential-families}, if $\vec{\eta}$ is the identity, and $X$'s transition matrix is more generally the \abbr{CEF} given in \cref{eq:def-cef}, then
$
	\E_{\vec{\theta}}[\vec{\tau}(X_t, X_{t+1}) \given X_0, \dotsc, X_t] = \nabla \lnpart(X_t, \vec{\theta})
$
for any $t \in \N$.
If $\vec{\eta}$ is not the identity function we apply the chain rule, replacing $\nabla \lnpart(X_t, \vec{\theta})$ with $J^{-1}_{\vec{\theta}} \nabla \lnpart(X_t, \vec{\theta})$, as in \textcite[Problem 5.6(b), \pnfmt{66}]{lehmann-casella-point-est}.
However, because $X$'s transition matrix is the \abbr{MEF} \cref{eq:def-mef}, $\lnpart$ is constant with respect to $X_t$, so we must replace $J^{-1}_{\vec{\theta}} \nabla \lnpart(X_t, \vec{\theta})$ with $J^{-1}_{\vec{\theta}} \nabla \lnpart(\vec{\theta})$.
Taking expectations on both sides we get, for any $t \in \N$,
\[
	J^{-1}_{\vec{\theta}} \nabla \lnpart(\vec{\theta})
	= \E_{\vec{\theta}}\big(\E_{\vec{\theta}}[\vec{\tau}(X_t, X_{t+1}) \given X_0, \dotsc, X_t]\big)
	= \E_{\vec{\theta}}[\vec{\tau}(X_t, X_{t+1})].
	\qedhere
\]

		\end{proof}

\section{Permutation-Uniform Markov Chains}\label{sec:puniformity}
	In this \lcnamecref{sec:puniformity}, we show that if every row of a transition matrix is a permutation of the other rows, then we may identify the corresponding Markov chain with an \abbr{IID} sequence on the same state space.
	This identification in \cref{thm:uniform-chain}, the main \lcnamecref{thm:uniform-chain} of the \lcnamecref{sec:puniformity}, preserves the exponential family representation of the transition matrix and the interpretability of the sufficient statistic when it is transformed into the joint distribution for the \abbr{IID} sequence.
	Autoregressive processes on discrete state spaces and several of the examples in \cref{sec:examples-lit} provide applications for this technique.

	The main concept in this \lcnamecref{sec:puniformity} is \emph{permutation uniformity}, the property that ``every row is a permutation of every other row.''
	For example, $\left[\begin{smallmatrix}\theta & 1 - \theta \\ 1 - \theta & \theta\end{smallmatrix}\right]$ and $\left[\begin{smallmatrix}1 & 0 \\ 1 & 0\end{smallmatrix}\right]$ are permutation uniform but $\left[\begin{smallmatrix}1 & 2 \\ 3 & 4\end{smallmatrix}\right]$ is not.
	We make this rough definition precise as follows.
	Let $\mathscr{A}$ be any set.
	For each $a \in \St$, define the $a$th \introduce{row} of a function or matrix $f \colon \St \times \St \to \mathscr{A}$ to be the function or row vector $b \mapsto f(a, b)$.
	We write permutations on $\St$ (i.e., bijections $\St \to \St$) juxtaposed with other permutations on $\St$ to denote composition and juxtaposed with elements of $\St$ to denote application.
	\begin{definition}\label{def:p-uniformity}
		Let $\perm \coloneqq \{\perm_a\}_{a \in \St}$ be a set of $|\St|$ permutations on $\St$ out of the $|\St|!$ possible.
		We say that $f$ is \introduce{permutation uniform}, or \introduce{p-uniform}, \introduce{under} $\perm$ if $f(a, \perm^{-1}_a c) = f(b, \perm^{-1}_b c)$ for all $a, b, c \in \St$.
		We drop the ``under $\perm$'' part to assert the existence of some permutations under which $f$ is p-uniform.
	\end{definition}
	A Markov chain is a \introduce{permutation-uniform Markov chain}, or \introduce{p-uniform chain}, if its transition matrix is p-uniform.
	We derived our definition from \textcite[\S~3]{rosenblatt-stationary-proc-as-shifts}; a reconciliation of that one with \textcite[Def.~1.4]{markov-chains-as-random-walks} appears in \cref{sec:appendx-yanirosenblatt}.

	While our focus is on the statistics of p-uniform chains, previous investigations of them have largely occurred in the language of dynamic systems.
	One way this bears out is that some authors have used different terminology for p-uniform chains, such as \emph{$\rho$-uniform stochastic graphs} in \textcite{Rubshtein:2004tx}.
	Another way is that results on p-uniformity have focused on Markov chains indexed by $\Z$ rather than $\N$, meaning that the chains have no initial states.

	The goal of several of these studies—and of the next \lcnamecref{sec:puniform-indep}—has been to connect a p-uniform chain $X$ with some \abbr{IID} sequence $Z$.
	\Textcite[Lem.~3]{rosenblatt-stationary-proc-as-shifts} showed on a countable state space that when $X$'s transition matrix $P$ has distinct entries in a row, $X_i$ and $X_{i+1}$ almost surely uniquely determine the \abbr{IID} $Z_i$s, each of which are independent of $X_{i-1}, X_{i-2}, \dotsc$.
	\Textcite{Rosenblatt:1960uw} constructed $Z_i$ \abbr{IID} uniformly on $[0, 1]$ such that $X_i$ is a function of all the preceding $Z_i$s.
	\Textcite{hanson-1963} extended this result to continuous state spaces.
	The proof of \textcite[Lem.~3]{rosenblatt-stationary-proc-as-shifts} spells out how to construct the $Z_i$s, but the distinctness constraint on the entries of $P$ limits the applicability of the lemma.
	\Textcite{Rosenblatt:1960uw} and \textcite{hanson-1963} relied on the Markov chain's having no beginning to prove the existence of $Z$ without constructing it (e.g., the former used the Borel-Cantelli lemma).
	In all three models, the $Z_i$s take values in a different space from the $X_i$s, or at least are not surjective onto the state space of the $X_i$s.

	\subsection{Independence}\label{sec:puniform-indep}
		In this \lcnamecref{sec:puniform-indep}, we derive \abbr{IID} $Z_i$s uniquely determined by a p-uniform Markov chain $X$ with transition matrix $P$.
		In contrast to prior studies of p-uniform chains, the entries of a row of $P$ need not be distinct, the $Z_i$s take on exactly the same values as the $X_i$s, and $X$ does not go infinitely into the past, but rather starts at time zero with $X_0, X_1, \dotsc$.

		The $Z_i$s' distribution can stand in for the $X_i$s', thus making statistical inference on the time-dependent $X_i$s easier by using the independent $Z_i$s.
		This makes the Markov chain $X$ rather like a random walk.
		We explore this connection more after the main theorem.

		$X$'s being p-uniform imposes enough structure for us to observe the desired \abbr{IID} sequence $Z$ from $X$, which we mean in the following sense.
We say that the $\St$-valued sequence of random variables $Z_1, \dotsc, Z_t$, $t \in \N_{>0}$, has a joint distribution \introduce{similar} under $\Pr$ to that of the $\St$-valued sequence of random variables $X_0, X_1, \dotsc, X_t$'s for some probability measure $\Pr$ if, for all $x_0, x_1, \dotsc, x_t \in \St$, there exist $z_1, \dotsc, z_t \in \St$ such that
\[
	\Pr(X_t=x_t, \dotsc, X_1 = x_1 \given X_0 = x_0) = \Pr(Z_t = z_t, \dotsc, Z_1 = z_1).
\]
For $\{Z_t\}_{t=1}^\infty$ to be \abbr{IID} with a \introduce{common law} (whose \abbr{PMF} under $\Pr$ is) $\mu$ means that $\Pr(Z_t = z) = \mu(z)$ for all $t \in \N_{>0}$ and all $z \in \St$.

		Recall from \cref{def:p-uniformity} that $\perm = \{\perm_a\}_{a\in\St}$ is a set of permutations on $\St$.

		\begin{theorem}\label{thm:uniform-chain}
			$X$ is a Markov chain on $\St$ under probability measure $\Pr$ with transition matrix $P$ p-uniform under $\perm$ if and only if there exist a \abbr{PMF} $\mu$ on $\St$, a sequence of $\St$-valued random variables $Z \coloneqq \{Z_t\}_{t=1}^\infty$, and an $\St$-valued random variable $X_0$ such that
\begin{conditions}
	\item\label{itm:uniform-chain-common-law}
		$Z$ is \abbr{IID} with common law $\mu$;
	\item\label{itm:uniform-chain-x-z-eq}
		$X_{t+1} = \perm^{-1}_{X_t}Z_{t+1}$ $\Pr$-almost surely for all $t \in \N$;
	\item\label{itm:uniform-chain-p-uniform}
		$P(a, b) = \mu(\perm_a b)$ for all $a, b \in \St$; and
	\item\label{itm:uniform-chain-indep}
		$X_0, Z_1, \dotsc, Z_t$ are mutually independent for all $t \in \N_{>0}$.
\end{conditions}
When such a $Z$ exists, $X$ and $Z$ have similar joint distributions.
Further, for all $t \in \N$, $Z_{t+1}$ and any random vector of the form $(X_{i_1}, \dotsc, X_{i_n}, X_t)$ for $i_1, \dotsc, i_n \in \{0, 1, \dotsc, t-1\}$ and $n \in \{0, 1, \dotsc, t-1\}$ are pairwise independent.

		\end{theorem}

		If we are given a p-uniform Markov chain $X$ with transition matrix $P$, \cref{thm:uniform-chain-yano-definition} already tells us what $\mu$ has to be: any permutation of any row of $P$.
		We are free to choose any permutation of a row of $P$ that is convenient because we can choose any permutations $\perm$ that make $P(a, b) = \mu(\perm_a b)$ true.
		In this sense, once we're given $P$, we have no choice over $\mu$; we can only choose the permutations $\perm$.
		The choice of permutations $\perm$ determines the sequence $Z$ almost surely from $X$.
		Applying \cref{thm:uniform-chain} thus comes down to choosing $\perm$s.
		Conversely if we are given $Z$ with the common law $\mu$, different choices of $\perm$s give rise to different Markov chains $X$.
		\Vref{ex:density-stability} illustrates this phenomenon.

		Perhaps surprisingly, \cref{thm:uniform-chain} does not care about $X_0$'s distribution.
		However, $X_0$'s only role in the \lcnamecref{thm:uniform-chain} is to condition other distributions.

		\begin{proof}[Proof of \cref{thm:uniform-chain}]
			\textbf{Forward Implication.}
Suppose $X$ is a Markov chain on $\St$ under probability measure $\Pr$ with transition matrix $P$ p-uniform under $\perm$.

\Cref{thm:uniform-chain-yano-definition} provides for the existence of a \abbr{PMF} $\mu$ on $\St$ such that $P(a, b) = \mu(\perm_a b)$, \cref{itm:uniform-chain-p-uniform}.
$X_0$ is an $\St$-valued random variable because $X$ is a Markov chain.

We can define the $\St$-valued random variables $Z = \{Z_t\}_{t\in\N_{>0}}$ by
\[
	Z_{t+1} \coloneqq \perm_{X_t}X_{t+1} \qquad \text{for all } t \in \N.
\]
This equality holds everywhere in the sample space and $\perm_a$ is invertible for all $a \in \St$, so $X_{t+1} = \perm^{-1}_{X_t}Z_{t+1}$ holds $\Pr$-almost surely for all $t \in \N$, establishing \cref{itm:uniform-chain-x-z-eq}.
In the remainder of this direction of the proof, we will make use of the equivalence between $Z_t = z$ and $X_t = \perm^{-1}_{X_{t-1}}z$ when $t \in \N_{>0}$.

Now we must show that $Z$ is \abbr{IID} with common law $\mu$, which is \cref{itm:uniform-chain-common-law}.
The work to prove it will also prove \cref{itm:uniform-chain-indep}.
To do so, we fix an arbitrary sequence $\{z_t\}_{t\in\N_{>0}}\subseteq \St$ and time $T \in \N_{>0}$, and we show that
\begin{equation}\label{eq:uniform-chain-goal-iid}
	\Pr(Z_1 = z_1, \dotsc, Z_T = z_T) = \mu(z_1) \dotsm \mu(z_T).
\end{equation}

For each $t \in [T]$, $Z_t = z_t$ determines a transition.
Stringing together all $T$ transitions determines the state of the Markov chain through time $T$ if we know the starting position $X_0$.
This suggests starting the proof with the left-hand-side expression in \cref{eq:uniform-chain-goal-iid} and using the law of total probability:
\begin{equation}\label{eq:uniform-chain-law-of-total-probability}
	\Pr(Z_1 = z_1, \dotsc, Z_T = z_T)
	= \sum_{a \in \St}\Pr(Z_1 = z_1, \dotsc, Z_T = z_T \given X_0 = a)\Pr(X_0 = a).
\end{equation}

We can formalize what we meant by ``stringing together'' the transitions as follows.
For each $t \in \N$, let $x_t \colon \St \to \St$ be defined recursively by
\[
	x_t(a) \coloneqq
	\begin{cases}
		a                          & t = 0 \\
		\perm^{-1}_{x_{t-1}(a)}z_t & t > 0.
	\end{cases}
\]
For a given $a \in \St$, $\{x_t(a)\}_{t \in \N}$ is a deterministic sequence of elements of $\St$.
Let $a \in \St$ such that $\Pr(X_0 = a) > 0$.

As we now prove by induction, the event $A$ in which $Z_1 = z_1, \dotsc, Z_T = z_T$, and $X_0 = a$ is almost surely the same event as the event $B$ in which $X_0 = x_0(a), \dotsc, X_T = x_T(a)$.
Let $t \in [T]$.
First suppose we are in event $A$, so $Z_t = z_t$ and $X_0 = a = x_0(a)$.
Under the induction hypothesis that $X_{t-1} = x_{t-1}(a)$, we have $X_t = \perm^{-1}_{X_{t-1}}Z_t = \perm^{-1}_{x_{t-1}(a)}z_t = x_t(a)$.
Thus we are in event $B$ as well.
Second suppose we are in event $B$, so $X_0 = x_0(a) = a$, $X_{t-1} = x_{t-1}(a)$, and $X_t = x_t(a)$.
Then $
	Z_t
	= \perm_{X_{t-1}}X_t
	= \perm_{x_{t-1}(a)}x_t(a)
	= \perm_{x_{t-1}(a)}\perm^{-1}_{x_{t-1}(a)}z_t
	= z_t.
$
Thus we are in event $A$ as well.
Consequently, we may expand the left term in the summand of \cref{eq:uniform-chain-law-of-total-probability} using these equivalent events for $a \in \St$:
\begin{equation}\label{eq:uniform-chain-equivalent-events}
	\Pr(Z_1 = z_1, \dotsc, Z_T = z_T \given X_0 = a)
	= \Pr(X_1 = x_1(a), \dotsc, X_T = x_T(a) \given X_0 = a).
\end{equation}

$X$ is a Markov chain so we can expand the right side of \cref{eq:uniform-chain-equivalent-events} as
\begin{equation}\label{eq:uniform-chain-markov-property}
	\Pr(X_1 = x_1(a), \dotsc, X_T = x_T(a) \given X_0 = a)
	= \prod_{t=1}^T P\left(x_{t-1}(a), x_t(a)\right).
\end{equation}
Applying \cref{itm:uniform-chain-p-uniform}, we can then write for any $t \in [T]$
\begin{equation}\label{eq:uniform-chain-transition-probability}
	P\left(x_{t-1}(a), x_t(a)\right)
	= \mu(\perm_{x_{t-1}(a)}x_t(a))
	= \mu(\perm_{x_{t-1}(a)}\perm^{-1}_{x_{t-1}(a)}z_t)
	= \mu(z_t).
\end{equation}
Putting together \cref{eq:uniform-chain-equivalent-events,eq:uniform-chain-markov-property,eq:uniform-chain-transition-probability}, we get
\begin{equation}\label{eq:uniform-chain-conditionally-iid}
	\Pr(Z_1 = z_1, \dotsc, Z_T = z_T \given X_0 = a)
	= \prod_{t=1}^T \mu(z_t).
\end{equation}
This is \emph{almost} what we need.
Combining \cref{eq:uniform-chain-conditionally-iid,eq:uniform-chain-law-of-total-probability} yields our goal from \cref{eq:uniform-chain-goal-iid}:
\begin{align*}
	\Pr(Z_1 = z_1, \dotsc, Z_T = z_T)
	&= \smashoperator[l]{\sum_{a \in \St}}\Pr(Z_1 = z_1, \dotsc, Z_T = z_T \given X_0 = a)\Pr(X_0 = a) \\
	&= \smashoperator[l]{\sum_{a \in \St}}\left[\prod_{t=1}^T \mu(z_t)\right]\Pr(X_0 = a)
	=\! \prod_{t=1}^T \mu(z_t)\sum_{\mathclap{a \in \St}}\Pr(X_0 = a)
	= \prod_{t=1}^T \mu(z_t).
\end{align*}
This establishes \cref{itm:uniform-chain-common-law}.
Between \cref{eq:uniform-chain-conditionally-iid,eq:uniform-chain-goal-iid}, we see that $X_0, Z_1, \dotsc, Z_T$ are independent, establishing \cref{itm:uniform-chain-indep}.

\textbf{Backward Implication.}
We must show that $X = \{X_t\}_{t\in\N}$, defined in \cref{itm:uniform-chain-x-z-eq}, is a Markov chain with a transition matrix $P$, defined in \cref{itm:uniform-chain-p-uniform}, and that $P$ is p-uniform under $\perm$.
To do so, we fix an arbitrary time $t \in \N$ and vector $\vec{x} = (x_0, \dotsc, x_{t+1}) \in \St^{t+2}$ such that $\Pr(X_t = x_t) > 0$, and we show that
\begin{align}
	\label{eq:uniform-chain-goal-markov}
	\Pr(X_{t+1} = x_{t+1} \given X_0 = x_0, \dotsc, X_t = x_t)
	&= \mu(\perm_{x_t}x_{t+1}), \\
	\label{eq:uniform-chain-goal-uniform}
	\Pr(X_{t+1} = x_{t+1} \given X_t = x_t) &= \mu(\perm_{x_t}x_{t+1}).
\end{align}
Proving \cref{eq:uniform-chain-goal-markov,eq:uniform-chain-goal-uniform} together will show that $X$ has the Markov property, and proving \cref{eq:uniform-chain-goal-uniform} will show that $X$ is p-uniform by \cref{thm:uniform-chain-yano-definition,eq:uniform-chain}.
Since the time $t$ is arbitrary, proving \cref{eq:uniform-chain-goal-uniform} will also establish that $X$ is homogeneous, so that $P(a, b) = \mu(\perm_a b)$ is the transition matrix for $X$.
By \cref{thm:uniform-chain-yano-definition}, this will establish that $P$ is p-uniform under $\perm$.

As we now prove by induction, the event $C$ in which $X_0 = x_0, \dotsc, X_T = x_T$ (on which the left side of \cref{eq:uniform-chain-goal-markov} is conditioned) is, for all $T \in \N$, almost surely (a.s.) the same event as event $D$ in which $X_0 = x_0$ and $Z_1 = \perm_{x_0}x_1, \dotsc, Z_T = \perm_{x_{T-1}}x_T$.
We prove this for the case when $T > 0$.
Let $s \in [T]$.
First suppose we are in event $C$, so $X_{s-1} = x_{s-1}$, $X_s = x_s$, and $X_0 = x_0$.
We have $x_s = X_s = \perm^{-1}_{X_{s-1}}Z_s = \perm^{-1}_{x_{s-1}} Z_s$ a.s., so $Z_s = \perm_{x_{s-1}}x_s$ a.s.
Thus we are in event $D$ as well.
Second suppose we are in event $D$, so $Z_s = \perm_{x_{s-1}}x_s$, $Z_1 = \perm_{x_0}x_1$, and $X_0 = x_0$.
Under the induction hypothesis that $X_{s-1} = x_{s-1}$ a.s., we have $X_s = \perm^{-1}_{X_{s-1}}Z_s = \perm^{-1}_{x_{s-1}}\perm_{x_{s-1}}x_s = x_s$ a.s.
Thus we are in event $C$ as well.

We prove \cref{eq:uniform-chain-goal-markov} as follows.
The first equality uses $C = D$ a.s.~for $T = t$, and the second uses $Z$'s \abbr{PMF} $\mu$ from \cref{itm:uniform-chain-common-law} and the independence of $X_0,Z_1, \dotsc, Z_{t+1}$ from \cref{itm:uniform-chain-indep}.
\begin{multline*}
	\Pr(X_{t+1} = x_{t+1} \given X_0 = x_0, X_1 = x_1, \dotsc, X_t = x_t) \\
	= \Pr(Z_{t+1} = \perm_{x_t}x_{t+1} \given X_0 = x_0, Z_1 = \perm_{x_0}x_1, \dotsc, Z_t = \perm_{x_{t-1}}x_t)
	= \mu(\perm_{x_t}x_{t+1}).
\end{multline*}

We prove \cref{eq:uniform-chain-goal-uniform} as follows.
The first and third equalities below apply $C = D$ a.s.~for $T = t-1$, and the second uses \cref{itm:uniform-chain-common-law,itm:uniform-chain-indep}.
\begin{multline*}
	\Pr(X_{t+1} = x_{t+1}, X_t = x_t \given X_0 = x_0, X_1 = x_1, \dotsc, X_{t-1} = x_{t-1}) \\
	\begin{aligned}
		= \Pr(Z_{t+1} = \perm_{x_t} x_{t+1}, Z_t = \perm_{x_{t-1}}x_t \given X_0 = x_0, Z_1 = \perm_{x_0}x_1, \dotsc, Z_{t-1} = \perm_{x_{t-2}}x_{t-1}) & \\
		= \mu(\perm_{x_t} x_{t+1}) \Pr(Z_t = \perm_{x_{t-1}}x_t \given X_0 = x_0, Z_1 = \perm_{x_0}x_1, \dotsc, Z_{t-1} = \perm_{x_{t-2}}x_{t-1}) &
	\end{aligned}
	\\
	= \mu(\perm_{x_t} x_{t+1}) \Pr(X_t = x_t \given X_0 = x_0, X_1 = x_1, \dotsc, X_{t-1} = x_{t-1}).
\end{multline*}
Apply the law of total probability with fixed $x_t$ and $x_{t+1}$ and sum over $\vec{x} = (x_0, \dotsc, x_{t-1})$:
\begin{multline*}
	\Pr(X_{t+1} = x_{t+1}, X_t = x_t) \\
	\begin{aligned}
		= \sum_{\crampedclap{\vec{x} \in \St^t}} \Pr(X_{t+1} = x_{t+1}, X_t = x_t \given X_0 = x_0, \dotsc, X_{t-1} = x_{t-1})\Pr(X_0 = x_0, \dotsc, X_{t-1} = x_{t-1}) & \\
		= \mu(\perm_{x_t} x_{t+1}) \sum_{\crampedclap{\vec{x} \in \St^t}} \Pr(X_t = x_t \given X_0 = x_0, \dotsc, X_{t-1} = x_{t-1})\Pr(X_0 = x_0, \dotsc, X_{t-1} = x_{t-1}) &
	\end{aligned}
	\\
	= \mu(\perm_{x_t} x_{t+1}) \Pr(X_t = x_t).
\end{multline*}
Dividing both sides by $\Pr(X_t = x_t)$ yields \cref{eq:uniform-chain-goal-uniform}, as desired.

\textbf{Finishing the Proof.}
The fact that $C = D$ a.s.~proves $X$ and $Z$ have similar distributions.
Finally, to prove that $Z_{t+1}$ and $(X_{i_1}, \dotsc, X_{i_n}, X_t)$ are independent, let $z \in \St$ and observe that
\begin{align*}
	\Pr(Z_{t+1} = z &\given X_t = x_t, X_{i_n} = x_{i_n}, \dotsc, X_{i_1} = x_{i_1}) \\
	&= \Pr(X_{t+1} = \perm^{-1}_{x_t} z \given X_t = x_t)
	= \mu(\perm_{x_t}\perm^{-1}_{x_t} z)
	= \Pr(Z_{t+1} = z).
	\qedhere
\end{align*}

		\end{proof}

		\Cref{ex:modular-autoregression,ex:var} below translate into \cref{thm:uniform-chain}'s notation some random-walk-like models from the Markov-chain literature that turn out to be p-uniform.

		\begin{example}[Modular Autoregressive Model {\citereset\Parencite[Example 6.2, \pnfmt{66}]{diaconis-freedman-1999}}]\label{ex:modular-autoregression}
			\Citeauthor{diaconis-freedman-1999} proposed a model of Markov chains induced by iterating random functions.
			Details of how \cref{thm:uniform-chain} turns out to be a sub-model of their model are in \cref{sec:appendx-induced}, but the following \lcnamecref{ex:modular-autoregression} works equally well in the language of either model.

			Let $\St= \Z/n\Z$ be the set of $n$ integers modulo $n$.
			Fix some initial state $x_0 \in \St$ and set $X_0 \coloneqq x_0$.
			Then define $X_{t+1} \coloneqq X_t + Z_{t+1} \pmod{n}$, where the $Z_t$s are uniform, \abbr{IID} random variables taking values zero or one each with probability a half: $\mu(0) = \mu(1) = \slfrac{1}{2}$.
			Then $Z_{t+1} = X_{t+1} - X_t \pmod{n}$, and, since modular subtraction is bijective, $X$ is a p-uniform chain with $\perm_i j = j - i \pmod{n}$ for each $i, j \in \St$.
			The $i,j$ entry of the transition matrix is defined by $\mu(\perm_i j) = \mu(j - i \pmod{n})$, which is a half if $i - j \pmod{n}$ is either zero or one, and is zero otherwise.
			$X$ is irreducible and aperiodic, and it converges to its uniform stationary distribution at an exponential rate.
			\qed
		\end{example}

		\begin{example}[Vector Autoregressive Model {\citereset\Parencite[1171]{tensor-regression-panel-data}}]\label{ex:var}
			\Citeauthor*{tensor-regression-panel-data} introduced a multilinear tensor autoregressive model for network data similar to \abbr{VAR} models \autocite[\protect{\bibstring{confer}}][657]{wooldridge-ed5}.
			Let $\vec{X}_i$ be the vectorization of a network's weighted adjacency matrix at time $i$.
			The bilinear version of \citeauthor{tensor-regression-panel-data}'s model is
			\begin{align*}
				\vec{X}_i &= \theta \vec{X}_{i-1} + \vec{Z}_i, &
				\E(\vec{Z}_i) &= \vec{0}, &
				\E(\vec{Z}_i \vec{Z}^{\transp}_j) &=
				\begin{cases}
					\Sigma  & i = j \\
					\vec{0} & i \ne j,
				\end{cases}
			\end{align*}
			where $\theta$ and $\Sigma$ are matrices of parameters to be estimated.
			\Citeauthor*{tensor-regression-panel-data} used variations of this model on a time series of verbal and material diplomatic actions among 25 countries between 2004 and 2014.
			Geographically nearby countries' actions were the best predictors of each country's actions, with the exceptions of the United Kingdom and Australia and of the United States and certain other countries.
			The analysis also found that the relations between two countries depended on other countries' relations.

			With a couple simple restrictions, this model becomes a p-uniform Markov chain.
			First, we must restrict all values to rationals so that the state space for the $\vec{X}_i$s is countable.
			Second, we must assume that the $\vec{Z}_i$s are \abbr{IID}, which is compatible with the assumptions that $\E(\vec{Z}_i) = \vec{0}$ and $\E(\vec{Z}_i \vec{Z}^{\transp}_j) = \vec{0}$ if $i \ne j$.
			Then the set $\perm$ of permutations under which the $\vec{X}_i$s are p-uniform are those for which $\perm_{\vec{X}_{i-1}}\vec{X}_i = \vec{X}_i - \theta \vec{X}_{i-1} = \vec{Z}_i$.
		\qed
		\end{example}

	\subsection{Convergence}
		A statistical consequence of a Markov chain's p-uniformity is that we can apply \abbr{IID} convergence theorems to the Markov chain.
		In the case of p-uniform Markov chains that are \abbr{MEF}s, \cref{thm:puniform-as-convergence} strengthens results in the literature for \abbr{CEF}s where the function $\vec{\tau}$ in the theorem is a sufficient statistic.
		\Textcite[Prop.~1.1]{Stefanov:1995cl} proved convergence in probability of the time-average of an \abbr{MEF}'s sufficient statistic.
		\Cref{thm:puniform-as-convergence} strengthens this convergence to almost-sure and $L^1$ convergence.
		Coupling \cref{thm:puniform-as-convergence} with \cref{thm:mef-expected-suff-stat} gives a limit to the mean parameter of the \abbr{MEF}.
		\Textcite[Thm.~1, \pnfmt{598}]{conditional-exponential-families} had such a result for \abbr{CEF}s, but only in expectation conditional on the past state of the chain, with which condition \cref{thm:mef-expected-suff-stat} dispenses.

		\begin{theorem}\label{thm:puniform-as-convergence}
			Suppose $X$ is a Markov chain on $\St$ under $\Pr$ with transition matrix $P$ p-uniform under $\perm$.
			Let $\vec{\tau} \colon \St \times \St \to \R^\ell$ be p-uniform under $\perm$ such that $\sum_{b \in \St}\vec{\tau}(a, b)P(a, b)$ converges absolutely for all $a \in \St$.
			Then, for any $a \in \St$ and any $s \in \N$,
			\[
				\frac{1}{t} \sum_{i=0}^{t-1} \vec{\tau}(X_i, X_{i+1})
				\to
				\E_{\Pr}[\vec{\tau}(X_s, X_{s+1})]
				= \sum_{b \in \St}\vec{\tau}(a, b)P(a, b)
				\qquad
				\text{as } t \to \infty,
			\]
			where convergence is both almost sure and in $L^1$ under $\Pr$.
			In particular, the limit does not depend on $s$, the expectation does not depend on the initial distribution of $X_0$, and the equation does not depend on $a$.
		\end{theorem}

		\begin{proof}
			By \cref{thm:uniform-chain}, $Z_i \coloneqq \perm_{X_{i-1}}X_i$, $i \in \N_{>0}$, is a sequence of \abbr{IID}, $\St$-valued random variables with some common law $\mu$.
Since $\vec{\tau}$ is p-uniform under $\perm$, \cref{thm:uniform-chain-yano-definition} allows us to define $\vec{m} \colon \St \to \R^\ell$ by $\vec{m}(b) \coloneqq \vec{\tau}(a, \perm^{-1}_a b)$ for all $a, b \in \St$.

We claim that $\vec{m}(Z_1)$ has finite expectation.
Since $\sum_{b \in \St}\vec{\tau}(a, b)P(a, b)$ converges absolutely for all $a \in \St$, the Riemann rearrangement theorem says that every rearrangement converges absolutely to the same value \autocite[Thm.~3.55]{rudin}.
For each $a \in \St$, $\perm^{-1}_a$ is bijective, so
\begin{align}\label{eq:puniform-as-convergence-rearrangement}
	\sum_{b \in \St} \vec{\tau}(a, b)P(a, b)
	= \sum_{b \in \St} \vec{\tau}(a, \perm^{-1}_a b)P(a, \perm^{-1}_a b)
\end{align}
converges absolutely.
Further, \cref{thm:uniform-chain} says that $\mu$ satisfies $P(a, b) = \mu(\perm_a b)$ for all $b \in \St$, so
\begin{align}\label{eq:puniform-as-convergence-z-expectation}
	\sum_{b \in \St} \vec{\tau}(a, \perm^{-1}_a b)P(a, \perm^{-1}_a b)
	&= \sum_{b \in \St} \vec{m}(b)\mu(b)
	= \E_\mu[\vec{m}(Z_1)].
\end{align}
Combining \cref{eq:puniform-as-convergence-rearrangement,eq:puniform-as-convergence-z-expectation} yields that
\begin{equation}\label{eq:puniform-as-convergence-compute-expectation}
	\sum_{b \in \St} \vec{\tau}(a, b)P(a, b)
	= \E_\mu[\vec{m}(Z_1)]
\end{equation}
converges absolutely.
Thus $\vec{m}(Z_1)$ has finite expectation.

This allows us to apply Kolmogorov's strong law of large numbers \autocite[Thm.~20.2]{jp}:
\begin{equation}\label{eq:puniform-as-convergence-lln}
	\lim_{t \to\infty} \frac{1}{t}\sum_{i=0}^{t-1} \vec{m}(Z_{i+1})
	= \E_\mu[\vec{m}(Z_1)]
\end{equation}
holds both $\mu$-almost surely and in $L^1$.
Since the $Z_i$s have the law $\mu$ under $\Pr$, \cref{eq:puniform-as-convergence-lln} holds $\Pr$-almost surely.
For any $t \in \N_{>0}$, we have
\begin{align}\label{eq:puniform-as-convergence-averages}
	\frac{1}{t}\sum_{i=0}^{t-1} \vec{m}(Z_{i+1})
	&= \frac{1}{t}\sum_{i=0}^{t-1} \vec{\tau}(X_i, \perm^{-1}_{X_i} Z_{i+1})
	= \frac{1}{t}\sum_{i=0}^{t-1} \vec{\tau}(X_i, X_{i+1}).
\end{align}
For any $s \in \N$, the definition of expectation and $X$'s Markov property yield
\begin{align}
	\notag
	\E_{\Pr}[\vec{\tau}(X_s, X_{s+1})]
	&= \sum_{a \in \St} \sum_{b \in \St} \vec{\tau}(a, b)\Pr(X_s = a) P(a, b)
	= \sum_{a \in \St} \Pr(X_s = a) \sum_{b \in \St} \vec{\tau}(a, b)P(a, b), \\
	\intertext{and, by \cref{eq:puniform-as-convergence-compute-expectation},}
	\notag
	&= \sum_{a \in \St} \Pr(X_s = a) \E_\mu[\vec{m}(Z_1)] \\
	\label{eq:puniform-as-convergence-def-expectation}
	&= \E_\mu[\vec{m}(Z_1)].
\end{align}
Combining \cref{eq:puniform-as-convergence-compute-expectation,eq:puniform-as-convergence-lln,eq:puniform-as-convergence-averages,eq:puniform-as-convergence-def-expectation} yields the result.

		\end{proof}

	\subsection{Symmetry}\label{sec:symmetry}
		Sometimes we can deduce that a stationary distribution of a Markov chain p-uniform under $\perm$ is the uniform distribution just from the permutations $\perm$.
		We say that $\perm$ is \introduce{symmetric} if $\perm_a b = \perm_b a$ for all $a, b \in \St$.
		\Cref{thm:puniform-symmetric} carries the symmetry of $\perm$ through to the Markov chain's transition matrix.
		On a finite state space, a symmetric, stochastic matrix is doubly stochastic and thus has the uniform distribution as a stationary distribution; see \textcite{marshall-olkin}.

		\begin{lemma}\label{thm:puniform-symmetric}
			Let $P$ be an $\St \times \St$ matrix p-uniform under $\perm$, and let $\vec{\mu}$ be a vector such that $P(a,b) = \mu(\perm_a b)$.
			If $\perm$ is symmetric, then $P$ is symmetric.
			Conversely, if $P$ is symmetric and the entries of $\vec{\mu}$ are all distinct, then $\perm$ is symmetric.
		\end{lemma}

		A common set operator, which we need for \cref{sec:examples-lit}, exemplifies symmetry.

		\begin{example}\label{ex:puniform-symmetric-difference}
			All the elements in exactly one of the sets $A$ or $B$ constitute their \introduce{symmetric difference} $A \triangle B$.
			If the state space $\St$ is a field of sets (closed under intersections and unions), then we can define permutations $\perm = \{\perm_A\}_{A \in \St}$ on $\St$ by $\perm_A B \coloneqq A \triangle B$.
			$\triangle$'s commutativity makes $\perm$ typical of symmetric sets of permutations in that $\perm_\emptyset$ is the \emph{unique} identity permutation:
			If $\perm_A$ were also the identity, then $\perm_A A = A = \perm_\emptyset A$, and, by symmetry, $\perm_A A = \perm_A \emptyset$, so, by invertibility, $A = \emptyset$.
			\Cref{thm:puniform-symmetric} implies that any Markov chain p-uniform under the symmetric difference operator has the uniform distribution as a stationary distribution.
			\qed
		\end{example}

	\subsection{Permutation Uniformity and \abbr{CEF}s}\label{sec:puniformity-cefs}

		In this \lcnamecref{sec:puniformity-cefs}, we show that p-uniformity preserves \abbr{MEF} structure; we have already discussed convergence of mean parameters of p-uniform \abbr{MEF}s in and around \cref{thm:puniform-as-convergence}.

\Textcite{conditional-exponential-families} presented a theorem about \abbr{CAEF}s in a spirit similar to our present investigation of \abbr{CEF}s' relationship with \abbr{IID} sequences.
It supposed $X$ is real valued and $\mathscr{P}$ is a \abbr{CAEF} as in \cref{eq:def-cef} whose natural parameter space is open.
\Textcite[Thm.~3]{conditional-exponential-families} derived an additive process (partial sums of \abbr{IID} real-valued random variables) with the same law as a certain function of $X$ when $\vec{\tau}(a, \cdot)$ is invertible in the second slot.
On a finite state space and with $\kappa(a, \cdot)$ constant, invertibility of $\vec{\tau}(a, \cdot)$ in the second slot is equivalent to the transition matrix's having distinct numbers in every entry of a row.
This is similar to the distinctness requirement of \textcite[Lem.~3]{rosenblatt-stationary-proc-as-shifts}.
\Textcite[Thm.~4]{conditional-exponential-families} derived strong consistency of \abbr{MLE} and a central limit theorem for the Fisher information of a \abbr{CAEF} under the invertibility assumption.

In terms of \cref{thm:uniform-chain}, the next \lcnamecref{thm:uniform-cef-exp-fam} and its \lcnamecref{thm:uniform-cef-exp-fam-is-mef} show that if $X$ is p-uniform, then its distribution comes from an \abbr{MEF} if and only if $Z$'s distribution comes from an exponential family.
We say that $\mathscr{P}$ is \introduce{p-uniform} under $\perm$ if $P_{\vec{\theta}}(a, \perm^{-1}_a c) = P_{\vec{\theta}}(b, \perm^{-1}_b c)$ for all $a, b, c \in \St$ and all $\vec{\theta} \in \Theta$.
In this case, for each $\vec{\theta} \in \Theta$, \cref{thm:uniform-chain-yano-definition} says that there exists a \abbr{PMF} $\mu_{\vec{\theta}}$ such that $P_{\vec{\theta}}(a, b) = \mu_{\vec{\theta}}(\perm_a b)$ for all $a, b \in \St$.
Put $\mathscr{M} \coloneqq \{\mu_{\vec{\theta}}\}_{\vec{\theta} \in \Theta}$.

\begin{proposition}\label{thm:uniform-cef-exp-fam}
	Suppose $\mathscr{P}$ is p-uniform as above.
	If $\mathscr{P}$ is the \abbr{CEF} given in \cref{eq:def-cef}, then $\mathscr{M}$ is an exponential family:
	$
		\mu_{\vec{\theta}}(b)
		= \kappa(a, \perm^{-1}_a b)\exp\bigl(\vec{\eta}(\vec{\theta}) \cdot \vec{\tau}(a, \perm^{-1}_a b) - \lnpart(a, \vec{\theta})\bigr)
	$
	for any $a, b \in \St$ and $\vec{\theta} \in \Theta$.
	Conversely, if $\mathscr{M}$ is the exponential family given in \cref{eq:exp-fam}, then $\mathscr{P}$ is an \abbr{MEF}:
	$
		P_{\vec{\theta}}(a, b)
		= \kappa(\perm_a b) \exp\bigl(\vec{\eta}(\vec{\theta}) \cdot \vec{\tau}(\perm_a b) - \lnpart(\vec{\theta})\bigr)
	$
	for any $a, b \in \St$ and $\vec{\theta} \in \Theta$.
\end{proposition}

\begin{proof}
	Prove the forward implication by plugging $\mu_{\vec{\theta}}(b) = \mu_{\vec{\theta}}(\perm_a \perm^{-1}_a b) = P_{\vec{\theta}}(a, \perm^{-1}_a b)$ into \cref{eq:def-cef}.
	Prove the converse by plugging $P_{\vec{\theta}}(a, b) = \mu_{\vec{\theta}}(\perm_a b)$ into \cref{eq:exp-fam}.
\end{proof}

\begin{corollary}\label{thm:uniform-cef-exp-fam-is-mef}
	If $\mathscr{P}$ is p-uniform and a \abbr{CEF}, then it is an \abbr{MEF}.
\end{corollary}

\begin{proof}
	\Cref{thm:uniform-cef-exp-fam} says that, if $\mathscr{P}$ is p-uniform and a \abbr{CEF}, then $\mathscr{M}$ is an exponential family, which in turn implies that $\mathscr{P}$ is an \abbr{MEF}.
\end{proof}

Next, \cref{thm:puniform-suff-stat-implies-puniform-transition-matrix,thm:puniform-mef-implies-puniform-suff-stat} give necessary and sufficient conditions for an \abbr{MEF} to be p-uniform in terms of the p-uniformity of the carrier measure $\kappa$ and sufficient statistic $\vec{\tau}$.
First we give the sufficient condition%
,
		the proof of which is in \cref{proof:puniform-suff-stat-implies-puniform-transition-matrix}.%
		\begin{observation}\label{thm:puniform-suff-stat-implies-puniform-transition-matrix}
	If $\mathscr{P}$ is the \abbr{CEF} given in \cref{eq:def-cef} and $\vec{\tau}$ and $\kappa$ are p-uniform both under $\perm$, then $\mathscr{P}$ is p-uniform under $\perm$.
\end{observation}

		The exact converse of the above \lcnamecref{thm:puniform-suff-stat-implies-puniform-transition-matrix} is not quite true.
If $\kappa$ is zero in a couple positions, then the corresponding values of $\vec{\tau}$ are unconstrained.
We also need $\Theta$ to be big enough to determine which hyperplanes the different values of $\vec{\tau}$ lie in.
We manage the latter concern in \cref{thm:puniform-mef-implies-puniform-suff-stat} by assuming that $\vec{\eta}$ has
		\introduce{affinely independent entries}, meaning that $\vec{\delta} \cdot \vec{\eta}(\vec{\theta}) = h$ for all $\vec{\theta} \in \Theta$ implies $\vec{\delta} = \vec{0}$ and $h = 0$ \autocites[38]{Kuchler:1997vo}[40]{wainwright-jordan}.%
		\begin{theorem}\label{thm:puniform-mef-implies-puniform-suff-stat}
	Suppose that $\mathscr{P}$ is the \abbr{MEF} from \cref{eq:def-mef} and is p-uniform under $\perm$.
	Then $\kappa$ is p-uniform under $\perm$.
	Further, if $\kappa$ is never zero and $\vec{\eta}$ has affinely independent entries, then $\vec{\tau}$ is p-uniform under $\perm$.
\end{theorem}

\begin{proof}
	Fix $a, b, c \in \St$.
	If $\kappa(a, \perm^{-1}_a b) = 0$, then $0 = P_{\vec{\theta}}(a, \perm^{-1}_a b) = P_{\vec{\theta}}(c, \perm^{-1}_c b)$ for all $\vec{\theta} \in \Theta$.
	Since $\exp > 0$ on $\R$, we must have $\kappa(c, \perm^{-1}_c b) = 0$ as well.

	By the same token, suppose $\kappa(a, \perm^{-1}_a b) > 0$, and notice that $\kappa(c, \perm^{-1}_c b) > 0$ as well.
	Then rearranging the equation $P_{\vec{\theta}}(a, \perm^{-1}_a b) = P_{\vec{\theta}}(c, \perm^{-1}_c b)$ yields
	\begin{equation*}
		\vec{\eta}(\vec{\theta}) \cdot [\vec{\tau}(c, \perm^{-1}_c b) - \vec{\tau}(a, \perm^{-1}_a b)]
		= \log\frac{\kappa(a, \perm^{-1}_a b)}{\kappa(c, \perm^{-1}_c b)}
		\qquad \text{for all } \vec{\theta} \in \Theta.
	\end{equation*}
	Since the right side is constant with respect to $\vec{\theta}$, affine independence of $\vec{\eta}$'s entries implies that $\vec{\tau}(c, \perm^{-1}_c b) - \vec{\tau}(a, \perm^{-1}_a b) = \vec{0}$ and $\log(\slfrac{\kappa(a, \perm^{-1}_a b)}{\kappa(c, \perm^{-1}_c b)}) = 0$.
\end{proof}

A variety of hypotheses familiar in exponential family theory imply that $\vec{\eta}$ has affinely independent entries, justifying the application of \cref{thm:puniform-mef-implies-puniform-suff-stat}.
These hypotheses assume alternately that
\begin{enumerate*}
	\item
		\cref{eq:def-mef} is a \introduce{minimal representation} of the $a$th row of $P \in \mathscr{P}$, meaning that $\vec{\tau}(a, \cdot)$ and $\vec{\eta}$ both have affinely independent entries \autocites[Cor.~8.1, \pnfmt{113}]{barndorff}[38]{Kuchler:1997vo}[40]{wainwright-jordan};
	\item\label{itm:param-space-contains-open-set}
		$\vec{\eta}(\Theta)$ contains an open, $\ell$\hyphen dimensional set \autocite[Thms.~5.2.11 or 6.2.25, \pnfmt{217,288}]{casella-stat-infer}; or
	\item\label{itm:param-space-contains-indep-vecs}
		$\vec{\eta}(\Theta)$ contains $\ell + 1$ affinely independent vectors \autocite[Cor.~6.16, \pnfmt{39}]{lehmann-casella-point-est}.
\end{enumerate*}
\Cref{itm:param-space-contains-open-set} implies \cref{itm:param-space-contains-indep-vecs}, which in turn implies that $\vec{\eta}$ has affinely independent entries; the straightforward proof is
		in \cref{proof:param-space}.

\section{Markov Chains of Graphs}\label{sec:interp}
	In this \lcnamecref{sec:interp}, we apply the theories of \abbr{CEF}s and of p-uniform Markov chains to some of the network models that \textcite{hfx2010} proposed.
We find that some of them are p-uniform and thus \abbr{MEF}s.
For two of the models, we can avoid \abbr{MCMC} or Newton's method for \abbr{MLE}, which is what \citeauthor*{hfx2010} used; instead we give a closed form for the \abbr{MLE}.
We also explore the relationships among p-uniformity, \abbr{MEF}s, and statistical independence of the random edges in the graphs.

The main result in \cref{thm:suff-stat-dyadic-indep-multigraphs,thm:suff-stat-dyadic-indep-multigraph-joint-prob} is that we may replace $t$ observations of a p-uniform Markov chain of graphs with a single observation of a corresponding multigraph.
For this purpose, we narrow the scope of the exponential random multigraph model (\abbr{ERMM}) of \textcite{shafie2015} by introducing \emph{exponential random $t$-multigraph model}s, which differ from \abbr{ERMM}s by capping edge multiplicities at $t$.

We consider the state space $\St$ to be the set $\Graphs[t]{n}$ of \emph{loopless multigraphs}, notation for which we build up as follows.
We fix a number $n ≥ 2$ of \introduce{vertices} or \introduce{nodes} $[n] \coloneqq \{1, \dotsc, n\}$.
Each potential edge comes from the set $\Dyads{n} \coloneqq \binom{[n]}{2}$ of \introduce{dyads} whose elements $\{i, j\} \subseteq [n]$, $i ≠ j$, we may write as $ij$ or $ji$ when the meaning is clear.
That we do not allow edges from a node to itself makes the multigraphs in $\Graphs[t]{n}$ \introduce{loopless}.
Let $N \coloneqq \left|\Dyads{n}\right| = \binom{n}{2}$.
We fix a \introduce{maximum multiplicity} $t \in \N_{>0}$.
A \introduce{multigraph} $g \in \Graphs[t]{n}$ associates the vertex set $[n]$ with the \introduce{multiplicity} $g(f) \in \{0, 1, \dotsc, t\}$ of copies of each dyad $f \in \Dyads{n}$.
When $g(f) > 0$, we say that $f$ is an \introduce{edge} of $g$.

We identify a multigraph $g$ with its \introduce{edge-multiplicity vector} $\vec{g} \in \{0, 1, \dotsc, t\}^{\Dyads{n}} \equiv \Graphs[t]{n}$.
This is the vectorization of the adjacency matrix of $g$.
The \introduce{edge set} $E(\vec{g})$ is $\{f \in \Dyads{n} \given g(f) > 0\}$.
The \introduce{complement} of $\vec{g}$ is $\overline{\vec{g}} \coloneqq t \vec{1} - \vec{g}$.

An \introduce{exponential random graph model} (\abbr{ERGM}) is an exponential family defined on simple graphs, $\Graphs[1]{n}$.
We introduce the analogous family of models for multigraphs.
\begin{definition}
	We call an exponential family defined on $\Graphs[t]{n}$ an \introduce{exponential random $t$-multigraph model} ($t$-\abbr{ERMGM}).
\end{definition}

The choice of which \abbr{ERGM} to use in practice depends on identifying the sufficient statistics appropriate for specific data.
Those statistics may incorporate node covariates, leading to models that \textcite{Fienberg:1985cv} introduced and whose \abbr{MLE} \textcite{yan-jiang-fienberg-2018} investigated.
\Citeauthor*{Fienberg:1985cv} contrasted \introduce{microanalytic} studies, such as \textcite{Snijders:2001km}, employing node covariates with \introduce{macroanalytic} studies solely of network topology—our focus in the sequel.
Many popular macroanalytic models, which \textcite{goldenberg2010} surveyed, rely on statistics built on subgraph counts.
The simplest choice of subgraph is the single edge.

\begin{example}[Erdős-Rényi Graph Model]\label{ex:ergm}
	This \abbr{ERGM} arises by choosing edges of random graph $\vec{G}$ independently each with probability $p$.
	The probability of $\vec{G}=\vec{g}$ is $p^{|E(\vec{g})|}(1-p)^{N-{|E(\vec{g})|}} = \exp\left[\log(\slfrac{p}{(1-p)}) |E(\vec{g})| + N\log(1-p)\right]$.
	The parameter function is $\eta(p) = \log\frac{p}{1-p}$, sufficient statistic is $|E(\vec{g})|$, and the log-partition function is $-N\log(1-p)$ \autocite[\S~2.2]{chatterjee-diaconis-estimating-ergms-published}.
	\qed
\end{example}

The literature on network statistics has studied a variety of other subgraph counts from a variety of viewpoints.
\Textcite{chatterjee-diaconis-estimating-ergms-published} treated subgraph counts in a general context.
Specific choices of subgraphs were made in the studies in \textcite{ergms-hard} of counts of triangles, in \textcite{park-newman-2-star} of counts of two-stars, or in \textcite{Snijders:2006gk} of a complicated combination of degree counts, triangles, and two-stars.
\textcites{Holland:1981fk}{chatterjee-diaconis-degree-sequence}[\S~2]{mle-beta-model} used degree sequences---equivalent to counting labeled $k$-stars.
The latter's generalized $\degseq$ model is equivalent to the degree-sequence model for multigraphs of \textcite{frankshafie2018}.

\begin{example}[$\degseq$ Model]\label{ex:beta-model}
	This is the \abbr{ERGM} whose sufficient statistic is the degree sequence $\degseq(\vec{G})$.
	The probability of $\vec{G} = \vec{g}$ is
	$
		\exp\left(\degseq(\vec{g}) \cdot \log\vec{\theta} - \lnpart(\vec{\theta})\right)
		= e^{-\lnpart(\vec{\theta})}\prod_{uv \in E(\vec{g})}\theta_u \theta_v.
	$
	The parameter $\theta_v$ represents the attractiveness of vertex $v$ \autocite{petrovic2016}.
	The log-partition function is $\lnpart(\vec{\theta}) = \prod_{u=1}^n \prod_{v=1}^{u-1}(\theta_u \theta_v + 1)$ \autocite[\S~1.2]{chatterjee-diaconis-degree-sequence}.
	\Textcite[Thm.~1.5]{chatterjee-diaconis-degree-sequence} gave an algorithm for the \abbr{MLE} of $\vec{\theta}$.
	\qed
\end{example}
\Cref{sec:loyalty-tergm} introduces a dynamic relative of this $\degseq$ model.

	\subsection{Finite Exchangeability}\label{sec:interp-exch}
		Two multigraphs $\vec{b}, \vec{c} \in \Graphs[t]{n}$ are \introduce{isomorphic}, written $\vec{b} \sim \vec{c}$, if there exists a bijection $\phi$ on $[n]$ such that $b(\phi(i)\phi(j)) = c(\phi(i)\phi(j))$ for all $i, j \in [n]$.
		Isomorphism is an equivalence relation.
		If $\mu$ is any \abbr{PMF} (or any other function) on $\Graphs[t]{n}$, then $\mu$ is \introduce{finitely exchangeable} if
		\begin{equation}\label{eq:p-uniform-equiv-rel-g-const}
			\vec{b} \sim \vec{c} \implies \mu(\vec{b}) = \mu(\vec{c}).
		\end{equation}
		\Textcite{lauritzen-graphical-exch,lauritzen-exch-networks2018} defined and analyzed finitely exchangeable \abbr{PMF}s on $\Graphs[1]{n}$.
		The former showed that the set of all such \abbr{PMF}s on $\Graphs[1]{n}$ form an exponential family whose sufficient statistic counts subgraphs of the random network by isomorphism class.
		The latter related the finitely exchangeable distributions of random networks to the marginal distributions of their subgraphs, and gave a de Finetti-like theorem for representing those distributions.

		We can use \cref{thm:p-uniform-exchangeable-networks} to extend finite exchangeability from (or to) a transition matrix $P$ on $\Graphs[t]{n} \times \Graphs[t]{n}$ p-uniform under $\perm$ to (or from) a \abbr{PMF} $\mu$ on $\Graphs[t]{n}$ such that $P(\vec{a}, \vec{b}) = \mu(\perm_{\vec{a}} \vec{b})$.
		To do so requires of $\perm$ that $\sim$ be \introduce{invariant under} $\perm$ in the sense that
		\begin{equation}\label{eq:p-uniform-equiv-rel-sigma-forward}
			\vec{b} \sim \vec{c} \implies \perm_{\vec{a}} \vec{b} \sim \perm_{\vec{a}} \vec{c} \text{ for all } \vec{a} \in \Graphs[t]{n}.
		\end{equation}
		Several simple properties of invariance under $\perm$ for any equivalence relation (not just isomorphism) are straightforward to show on any state space (not just $\Graphs[t]{n}$).
		\Cref{eq:p-uniform-equiv-rel-g-const}'s converse and
		\begin{equation}\label{eq:p-uniform-equiv-rel-f-const}
			\vec{b} \sim \vec{c} \implies P(\vec{a}, \vec{b}) = P(\vec{a}, \vec{c}) \text{ for all } \vec{a} \in \Graphs[t]{n}
		\end{equation}
		together imply $\sim$ is invariant under $\perm$; and
		\cref{eq:p-uniform-equiv-rel-f-const}'s converse and \cref{eq:p-uniform-equiv-rel-g-const} together imply $\sim$ is invariant under $\perm^{-1} \coloneqq \{\perm^{-1}_{\vec{a}}\}_{\vec{a} \in \Graphs[t]{n}}$.
		($\sim$'s invariance under $\perm^{-1}$ is just the converse of \cref{eq:p-uniform-equiv-rel-sigma-forward}, and vice versa.)
		However, since $\Graphs[t]{n}$ is finite (or, even if it weren't but if $\perm$ were closed under inversion), $\sim$'s invariance under $\perm$ is equivalent to $\sim$'s invariance under $\perm^{-1}$.

		\begin{proposition}\label{thm:p-uniform-exchangeable-networks}
			On $\Graphs[t]{n}$, suppose we have a transition matrix $P$, permutations $\perm$, and a \abbr{PMF} $\mu$ such that $P(\vec{a}, \vec{b}) = \mu(\perm_{\vec{a}} \vec{b})$ for all $\vec{a}, \vec{b} \in \Graphs[t]{n}$ and such that $\sim$ is invariant under $\perm$.
			Then $\mu$ is finitely exchangeable if and only if every row of $P$ is (i.e., \cref{eq:p-uniform-equiv-rel-f-const} holds), and every row of $P$ is finitely exchangeable if and only if some row of $P$ is.
		\end{proposition}

		\begin{proof}[Proof sketch.]
			\Cref{eq:p-uniform-equiv-rel-f-const} and $\sim$'s invariance under $\perm^{-1}$ together imply \cref{eq:p-uniform-equiv-rel-g-const};
			\cref{eq:p-uniform-equiv-rel-g-const} and $\sim$'s invariance under $\perm$ together imply \cref{eq:p-uniform-equiv-rel-f-const}.
			If there is an $\vec{a} \in \Graphs[t]{n}$ for which $\vec{b} \sim \vec{c} \implies P(\vec{a}, \vec{b}) = P(\vec{a}, \vec{c})$, then $\sim$'s invariance under both $\perm$ and $\perm^{-1}$ implies \cref{eq:p-uniform-equiv-rel-f-const} holds.
		\end{proof}

		One permutation set $\perm$ preserving isomorphism classes is the multigraph-complement operation $\vec{b} \mapsto \overline{\vec{b}}$.
		That is, $\sim$ is invariant under $\perm$ when $\perm_{\vec{a}} \vec{b} = \overline{\vec{b}}$ for all $\vec{a}, \vec{b} \in \Graphs[t]{n}$.
		Finitely exchangeable distributions include any in exponential families for which the carrier measure and the sufficient statistic are finitely exchangeable.
		Edge counts (\cref{ex:ergm}) are constant on isomorphism classes.
		Degree sequence (\cref{ex:beta-model}) is not constant on isomorphism classes, but degree distribution, a contingency table of degree counts \autocite[\S~5]{lauritzen-graphical-exch}, the degree sequence after sorting, or any other of unlabeled subgraphs are.

	\subsection{Dyadic Independence}\label{sec:model-di}
		Suppose $\vec{G}$ is a $\Graphs[t]{n}$-valued random variable.
If $\{G(f) \given f \in \Dyads{n}\}$ are mutually independent, then $\vec{G}$ (or equivalently, its distribution) is \introduce{dyadically independent} \autocite[``Methods'']{goodreau2007}.
Imposing dyadic independence on a model can be appropriate in \textcquote[1039]{graham-2017}[.]{settings where the drivers of link formation are predominately bilateral in nature, as may be true in some types of friendship and trade networks as well as in models of (some types of) conflict between nation-states}.

The remainder of this \lcnamecref{sec:model-di} describes conditions under which $\vec{G}$, the $\Graphs[t]{n}$-valued random variable, is dyadically independent and how dyadic independence interacts with p-uniformity.
\Cref{sec:di-ermgm} extends some definitions and basic facts about dyadic independence from graphs to multigraphs.
\Cref{sec:model-di-iid-multigraphs}'s main result, \cref{thm:suff-stat-dyadic-indep-multigraph-joint-prob}, allows us quickly to convert $t+1$ samples from a p-uniform \abbr{MEF} of simple graphs into a single observation of a $t$-\abbr{ERMGM}.

\subsubsection{Exponential Random Multigraph Models}\label{sec:di-ermgm}
	\cref{thm:factor-over-edges-dyadic-indep} characterizes a $t$-\abbr{ERMGM}'s dyadic independence in terms of its sufficient statistic and carrier measure.
	We say that a function $\vec{\tau} \colon \Graphs[t]{n} \to \R^\ell$ is \introduce{dyadditive}, or \introduce{factors over edges} \autocite[Eq.~2]{hfx2010}, if there are functions $\vec{\tau}_f \colon \{0, 1, \dotsc, t\} \to \R^\ell$ for each dyad $f \in \Dyads{n}$ such that
	\begin{equation}\label{eq:dyadditive}
		\vec{\tau}(\vec{g}) = \sum_{f \in \Dyads{n}}\vec{\tau}_f(g(f)).
	\end{equation}

	When $t = 1$ (the \abbr{ERGM} case), $\vec{\tau}$ is dyadditive if and only if there is a real, $\ell \times \Dyads{n}$ matrix $Q$ such that $Q \vec{g} = \vec{\tau}(\vec{g}) - \vec{\tau}(\vec{0})$; then the $f$th column of $Q$ is $\vec{\tau}_f(1) - \vec{\tau}_f(0)$.
	When the carrier measure is also constant, it is already known that an \abbr{ERGM} is dyadically independent if and only if its sufficient statistic is dyadditive \autocites[\S~3.2]{hunter2008goodness}[Eq.~20]{consistency-exp-fam-mle}.

	\begin{example}\label{ex:dyadditive}
		Let $\vec{g} \in \Graphs[1]{n}$.
		A sufficient statistic in the Erdős-Rényi graph model is the number $|E(\vec{g})|$ of edges.
		The number of edges is dyadditive because $|E(\vec{g})| = \vec{1} \cdot \vec{g}$.

		The $\degseq$ model's sufficient statistic, the degree sequence $\degseq(\vec{g})$, is dyadditive.
		To see this, let $K \in \{0, 1\}^{n \times \Dyads{n}}$ be the complete graph's \introduce{incidence matrix}, whose columns are the indicator $n$-vectors of the two-element sets constituting $\Dyads{n}$.
		Then $\degseq(\vec{g}) = K \vec{g}$.

		Statistics that rely on cliques larger than single edges are not dyadditive.
		\qed
	\end{example}

	When an \abbr{ERGM}'s carrier measure is not constant, \cref{thm:factor-over-edges-dyadic-indep} still specifies some cases where dyadditivity of the sufficient statistic implies dyadic independence.
	To describe those cases, we say that a function $\kappa \colon \Graphs[t]{n} \to \R$ is \introduce{dyadically multiplicative} if there are functions $\kappa_f \colon \{0, \dotsc, t\} \to \R$ for each dyad $f \in \Dyads{n}$ such that
	\begin{equation}\label{eq:dyadically-multiplicative}
		\kappa(\vec{g}) = \prod_{f \in \Dyads{n}}\kappa_f(g(f)).
	\end{equation}
	Dyadically multiplicative carrier measures in \abbr{ERGM}s include the common cases in which the carrier measure is constant.

	\begin{lemma}\label{thm:factor-over-edges-dyadic-indep}
		$\vec{G}$'s \abbr{PMF} is \cref{eq:exp-fam} such that $\vec{\tau}$ is dyadditive as in \cref{eq:dyadditive} and $\kappa$ is dyadically multiplicative as in \cref{eq:dyadically-multiplicative} if and only if $\vec{G}$ is dyadically independent with the \abbr{PMF} for multiplicity $G(f)$ being
		\begin{equation}\label{eq:ermgm-dyad}
			\newcommand{\expr}[1]{\kappa_f(#1)\exp\left(\vec{\eta}(\vec{\theta}) \cdot \vec{\tau}_f(#1)\right)}
			\mu^f_{\vec{\theta}}(m) = \frac{\expr{m}}{\sum_{r=0}^t\expr{r}},
			\qquad
			\begin{aligned}
				m &\in \{0, \dotsc, t\}, \\
				f &\in \Dyads{n}.
			\end{aligned}
		\end{equation}
	\end{lemma}

		The proof of the \lcnamecref{thm:factor-over-edges-dyadic-indep}, whose details we relegate to \cref{proof:factor-over-edges-dyadic-indep}, uses the law of total probability, \cref{sec:appendx-proofs}'s \cref{thm:sum-product-exchange}, and some combinatorics.

			Calculating the log-partition function, and thus \abbr{PMF}, of an \abbr{ERGM} is computationally intractable for large $n$: it is $\sharp\mathcal{P}$-hard and inapproximable in polynomial time \autocite{ergms-hard}.
	However, for the special case of dyadically independent \abbr{ERGM}s, computing the log-partition function requires a number of multiplications merely quadratic in $n$ \autocite[Example 1]{frank-strauss-1986}.
	The following corollary of \cref{thm:factor-over-edges-dyadic-indep}, which follows directly from expanding \cref{eq:factor-over-edges-dyadic-indep-denom}, generalizes this existing result to $t$-\abbr{ERMGM}s.

	\begin{observation}
		Under the conditions of \cref{thm:factor-over-edges-dyadic-indep}, the partition function is
		\[
			e^{\lnpart(\vec{\theta})}
			= \prod_{u=1}^n\prod_{v=1}^{u-1} \sum_{r=0}^t \kappa_{uv}(r)\exp\left(\vec{\eta}(\vec{\theta}) \cdot \vec{\tau}_{uv}(r)\right).
		\]
	\end{observation}

\subsubsection{Independent Sequences and Multigraphs}\label{sec:model-di-iid-multigraphs}
	Because of \cref{thm:uniform-chain}, we may be able to derive an \abbr{IID} sequence of simple graphs from a Markov chain of simple graphs.
	In this \lcnamecref{sec:model-di-iid-multigraphs}, we consider how to turn a sequence of $t$ \abbr{IID} draws from an \abbr{ERGM} into a single draw from a $t$-\abbr{ERMGM}.
	Define the \introduce{multigraph union} of $\vec{z}_1, \dotsc, \vec{z}_t \in \Graphs[s]{n}$ to be their vector sum $\vec{z}_1 + \dotsb + \vec{z}_t \in \Graphs[st]{n}$.
	Let $Z = (\vec{Z}_1, \dotsc, \vec{Z}_t)$ be an \abbr{IID} sequence of dyadically independent, $\Graphs[1]{n}$-valued random variables with multigraph union $\vec{W}$.
	Fix $z = (\vec{z}_1, \dotsc, \vec{z}_t) \in \Graphs[1]{n}^t$ with multigraph union $\vec{w}$.

	The following \lcnamecref{thm:suff-stat-dyadic-indep-multigraphs} says roughly that the order of the appearance of edges in $Z$ does not matter to $\vec{W}$.
		The proof is in \cref{proof:suff-stat-dyadic-indep-multigraphs}.
			\begin{theorem}[Dyadically Independent Multigraphs]\label{thm:suff-stat-dyadic-indep-multigraphs}
		With the notation above, $\vec{W}$ is dyadically independent and
		\begin{equation}\label{eq:suff-stat-dyadic-indep-multigraphs}
			\Pr(\vec{W} = \vec{w})
			= \Pr(Z = z)\prod_{f \in \Dyads{n}} \binom{t}{w(f)}.
		\end{equation}
	\end{theorem}

			We close this \lcnamecref{sec:model-di} by showing that taking multigraph unions preserves exponential family structure.
	The proof of the result is a straightforward combination of \textcite[Thm.~5.2.11]{casella-stat-infer} with \cref{eq:exp-fam,eq:suff-stat-dyadic-indep-multigraphs}, using the facts that $\vec{\tau}$ is dyadditive and that $\kappa$ is dyadically multiplicative.
	Notice that $Z$ does not appear in \cref{eq:suff-stat-dyadic-indep-multigraph-joint-prob}.

	\begin{theorem}\label{thm:suff-stat-dyadic-indep-multigraph-joint-prob}
		Suppose the common \abbr{PMF} of the components of $Z$ is $\mu_{\vec{\theta}}$ from \vref{eq:exp-fam} such that $\vec{\tau}$ is dyadditive as in \cref{eq:dyadditive} and $\kappa$ is dyadically multiplicative as in \cref{eq:dyadically-multiplicative}.
		Further, assume that $\vec{\eta}(\Theta)$ contains an open, $\ell$-dimensional set.
		Then the \abbr{PMF} of $\vec{W}$ also has an exponential family representation with the same parameter and parameter function, and with the sufficient statistic and carrier measure respectively equal to
		\begin{align}\label{eq:suff-stat-dyadic-indep-multigraph-joint-prob}
			\sum_{f \in \Dyads{n}} \vec{\tau}_f(1)W(f) - \vec{\tau}_f(0)\left[t - W(f)\right]
			\qquad \text{and} \qquad
			\prod_{f\in\Dyads{n}} \binom{t}{W(f)} \kappa_f(1)^{W(f)} \kappa_f(0)^{t - W(f)}.
		\end{align}
	\end{theorem}

	\subsection{Examples from the Literature}\label{sec:examples-lit}
		In this \lcnamecref{sec:examples-lit}, we run through some of the \abbr{TERGM} sufficient statistics that \textcite{hfx2010} proposed.
		Each is scalar, so $\ell = d = 1$ (\citeauthor*{hfx2010} coalesced them into a vector, but we examine each in isolation for simplicity).
		Throughout, $\St = \Graphs[1]{n}$, and $G = \{\vec{G}_i\}_{i \in \N}$ is a $\Graphs[1]{n}$-valued stochastic process that is a Markov chain under the \abbr{CEF} $\mathscr{P}$ of transition matrices $P_{\theta}$ from \vref{eq:def-cef} but with $\kappa \equiv 1$.
		Let $t \in \N_{>0}$.
		Notationally, $\vec{a}$, $\vec{b}$, and $\vec{c}$ are generic graphs.

		Scaling a sufficient statistic $\vec{\tau}$ by a constant $c$, often $c=n$ or $c=\frac{1}{n-1}$ so that $\vec{\tau}$ lies in $[0, n]$, occasionally improves interpretability while being transparent to the probability model because $(\frac{1}{c} \vec{\gamma}) \cdot (c \vec{\tau}) = \vec{\gamma} \cdot \vec{\tau}$.

		\begin{example}[Density and Stability {\citereset\Parencite[\S~2.1]{hfx2010}}]\label{ex:density-stability}
			If $\tau(\vec{a}, \vec{b}) = \frac{1}{n-1}\sum_{ij} b(ij)$, then $\tau$ is called \introduce{density}; if $\tau(\vec{a}, \vec{b}) = \frac{1}{n-1}\sum_{ij}b(ij)a(ij) + [1-b(ij)][1-a(ij)]$ then $\tau$ is called \introduce{stability}.
			Density measures the tendency of an edge to exist in the future irrespective of the present.
			Stability measures the tendency of an edge to continue existing or not in the future if it is doing so in the present.

			With p-uniformity as our hint, we notice that we can rewrite the statistics as the number of edges in bijective functions of the present and future networks.
			$\tau$ is density if $\tau(\vec{a}, \vec{b}) = |E(\vec{b})|$ or stability if $\tau(\vec{a}, \vec{b}) = \frac{1}{n-1}|E(\overline{\vec{a} \triangle \vec{b}})|$.
			Either way, $\tau$ is p-uniform under $\perm$:
			In the density case, $\perm_{\vec{a}}$ is the identity permutation for all $\vec{a}$; in the stability case, $\perm_{\vec{a}} \vec{b} = \overline{\vec{a} \triangle \vec{b}}$.
			$\mathscr{P}$ is p-uniform by \cref{thm:puniform-suff-stat-implies-puniform-transition-matrix}, and an \abbr{MEF} by \cref{thm:uniform-cef-exp-fam-is-mef}.
			By \cref{thm:uniform-chain}, in the density case, $\vec{Z}_i \coloneqq \vec{G}_i$ is an \abbr{IID} sequence of $\Graphs[1]{n}$-valued random variables; in the stability case, $\vec{Z}_{i} \coloneqq \overline{\vec{G}_{i-1} \triangle \vec{G}_{i}}$ is.

			In both cases, \cref{thm:uniform-cef-exp-fam} says that the common family $\{\mu_p\}_{p \in (0, 1)}$ of \abbr{PMF}s of the $\vec{Z}_i$s is an exponential family with the dyadditive sufficient statistic $\nu(\vec{b}) \coloneqq \tau(\vec{a}, \perm^{-1}_{\vec{a}} \vec{b}) = \frac{1}{n-1}|E(\vec{b})|$.
			Hence the $\vec{Z}_i$s are an \abbr{IID} sequence of Erdős-Rényi graphs (\cref{ex:ergm}) with canonical parameter $p$ and parameter function $\eta(p) = (n-1)\log\frac{p}{1-p}$.
			The $\vec{Z}_i$s are dyadically independent by \cref{thm:factor-over-edges-dyadic-indep}.

			\Cref{thm:puniform-as-convergence} says that, for any $j \in \N$, the mean parameter equals
			\[
				\lim_{t \to \infty} \frac{1}{t} \sum_{i=0}^t \tau(\vec{G}_i, \vec{G}_{i+1})
				= \E[\tau(\vec{G}_j, \vec{G}_{j+1})]
				= \sum_{\vec{b} \in \Graphs[1]{n}} \tau(\vec{0}, \vec{b}) P_p(\vec{0}, \vec{b})
				= p\frac{N}{n-1}.
			\]
			While \textcite{hfx2010} relied on \abbr{MCMC} \abbr{MLE} of parameters in the general case and Newton's method in the dyadditive case, in this particular case we compute the \abbr{MLE} $\hat{p}$ with a closed-form expression (see also \cref{thm:mef-expected-suff-stat}):
			\[
				\hat{p} = \frac{n-1}{tN} \sum_{i=0}^t \tau(\vec{G}_i, \vec{G}_{i+1}).
			\]

			Applying \cref{thm:suff-stat-dyadic-indep-multigraph-joint-prob}, let $\vec{W} \coloneqq \sum_{i=1}^t \vec{Z}_i$, which is a single, dyadically independent $t$-\abbr{ERMGM} random variable with sufficient statistic $\frac{1}{n-1} \vec{1} \cdot \vec{W}$.
			Then the joint distribution of $G$ is proportional, per \cref{thm:suff-stat-dyadic-indep-multigraphs}, to the distribution of $\vec{W}$, in which every edge is added to the multigraph by flipping a $p$-weighted coin for each dyad $t$ times.
			Put differently, for each dyad $f$, $W(f)$ is an independent, binomially distributed random variable with parameters $t$ and $p$.

			The main difference between density and stability is that, when $\tau$ is stability, $P_p$ is symmetric by \cref{thm:puniform-symmetric} because symmetric differences commute (see \cref{ex:puniform-symmetric-difference}).
			The unique stationary distribution is uniform because $\kappa \equiv 1$ implies that all entries of $P_p$ are positive.
			Linear algebraically, we can say more.
			Since $\overline{\vec{a} \triangle \vec{a}}$ is the complete graph with $N$ edges for any $\vec{a}$, the diagonal elements of $P_p$ are $P_p(\vec{a}, \vec{a}) = p^N$, and the trace of $P_p$ is $2^N p^N$.
			Every other off-diagonal entry of $P_p$ has a $(1-p)$ factor, so $P_p \to I$ as $p \to 1^-$ and $P_p \to \frac{1}{2^N-1}\left(\vec{1}\vec{1}^{\transp} - I\right)$ as $p \to 0^+$.
			At $p = \frac{1}{2}$, every row of $P_p = \frac{1}{2^N}\vec{1}\vec{1}^{\transp}$ is the uniform distribution.
			\qed
		\end{example}

		If $\tau$ is either of the statistics in the next two examples, then $\mathscr{P}$ is neither p-uniform nor an \abbr{MEF}.
		When $\mathscr{P}$ is not an \abbr{MEF}, \cref{thm:mef-exp-fam} says that $G$'s finite-sample joint distribution cannot have an exponential-family representation whose parameter is $\theta$.
		We are not guaranteed that $\sum_{i=0}^{t-1} \tau(\vec{G}_i, \vec{G}_{i+1})$ is sufficient for $\theta$.
		Even though stability and reciprocity (below) have similar looking formulae, the latter prevents analyzing $G$ as an exponential family in the same terms, $\theta$ and $\tau$, as its transition matrix was defined.
		We can interpret a density or stability \abbr{TERGM} either as an \abbr{IID} sequence of Erdős-Rényi graphs or as a single $t$-\abbr{ERMGM}, but such a simple interpretation of the statistics below is unavailable.

		\begin{example}[Reciprocity {\citereset\Parencite[\S~2.1]{hfx2010}}]
			Our focus has been on undirected graphs, but we briefly mention a statistic whose interpretation comes from its application to directed graphs, which we represent by ``vectors'' of adjacency matrices in $\{0, 1\}^{n \times n}$.
			Interpreting $\slfrac{0}{0}$ as zero, $\tau(\vec{a}, \vec{b}) \coloneqq n[\sum_{i,j = 1}^n a(i, j)]^{-1}\sum_{i,j = 1}^n b(j, i)a(i, j)$ is called \introduce{reciprocity}.

			To see that $\tau$ is not p-uniform, suppose that $\eta(\Theta)$ contains a number other than zero and that $n \ge 3$.
			Let $\vec{e}_1, \dotsc, \vec{e}_{n^2}$ be the standard basis of the set of adjacency matrices; these are the graphs containing exactly one edge.
			Then $\tau(\vec{e}_1, \vec{b}) \in \{0, n\}$ whereas $\tau(\vec{e}_1 \cup \vec{e}_2, \vec{e}_1) = \slfrac{n}{2}$.
			In fact, by violating the conclusion of \cref{thm:mef-uniqueness-of-suff-stats}, this also shows that $\mathscr{P}$ is not an \abbr{MEF}.
			\qed
		\end{example}

		\begin{example}[Transitivity {\citereset\Parencite[\S~2.1]{hfx2010}}]
			Interpreting $\slfrac{0}{0}$ as zero and taking sums over triples of vertices $i < j < k$, $\tau(\vec{a}, \vec{b}) \coloneqq n[\sum_{ijk} a(ij)a(jk)]^{-1} \sum_{ijk}b(ik)a(ij)a(jk)$ is called \introduce{transitivity}.
			Transitivity measures the tendency of edge $ik$ to come into existence in the future (graph $\vec{b}$) if edges $ij$ and $jk$ exist in the present (graph $\vec{a}$).
			That is the fraction of $\vec{a}$'s paths of length two (counted in the denominator) that try to close the triangle in graph $\vec{b}$.

			However, transitivity is not p-uniform and $\mathscr{P}$ is not an \abbr{MEF} when we suppose $n \ge 3$ and that $\eta(\Theta)$ contains a number other than zero.
			When $\vec{a}$ is a graph containing exactly two edges and those edges are adjacent, forming a path of length two, $\tau(\vec{a}, \vec{b})$ is either zero or $n$.
			Suppose $\vec{c}$ contains exactly one path of length three and no other edges; $\vec{c}$ then contains two paths of length two.
			If $\vec{b}$ contains an edge completing just one of those two triangles, then $\tau(\vec{c}, \vec{b}) = \slfrac{n}{2}$.
			Thus $\tau$ is not p-uniform, and, by \cref{thm:mef-uniqueness-of-suff-stats} and the fact that $\ell = 1$, $\mathscr{P}$ is not an \abbr{MEF}.
			Finally, when $\eta(\theta) \ne 0$, $P_{\theta}$ is not p-uniform: the $\vec{0}$th row of $P_{\theta}$ is the uniform distribution, but the $\vec{a}$th row for $\vec{a} \ne \vec{0}$ contains two different values.
			\qed
		\end{example}

	\subsection{A Model of Loyalty}\label{sec:loyalty-tergm}
		Motivated by the theoretical developments of the previous sections, we introduce a new \abbr{TERGM}.

\begin{example}[$\degseq$ \abbr{TERGM}]\label{ex:beta-tergm}
	By analogy with \vref{ex:beta-model}, we may define an \abbr{MEF} using the degree sequences of functions of the current and next graph.
	Using the same notation as the previous \lcnamecref{sec:examples-lit}, replace $d = \ell \coloneqq n$; $\Theta \coloneqq \R_{> \vec{0}}^n$, the positive orthant; $\vec{\eta} \coloneqq \log$; and $\vec{\tau}(\vec{a}, \vec{b}) \coloneqq \degseq(\perm_{\vec{a}} \vec{b})$, where $\degseq$ is the degree sequence.
	With these choices, $\mathscr{P}$ is an \abbr{MEF} p-uniform under any permutations $\perm$ on $\Graphs[1]{n}$.
	We call the model the \introduce{$\degseq$ \abbr{TERGM}} under $\perm$.

	For example, we may take use the symmetric difference operator as the permutations:
	\[
		P_{\vec{\theta}}(\vec{a}, \vec{b})
		= \exp\left(\degseq(\vec{a} \triangle \vec{b}) \cdot \log \vec{\theta} - \lnpart(\vec{\theta})\right)
		= \frac{\prod_{uv \in E(\vec{a} \triangle \vec{b})} \theta_u \theta_v}{\prod_{u=1}^n\prod_{v=1}^{u-1}(\theta_u \theta_v + 1)}.
	\]
	$\vec{Z}_i \coloneqq \vec{G}_{i-1} \triangle \vec{G}_i$ is an \abbr{IID} sequence of $\degseq$ \abbr{ERGM}s as in \cref{ex:beta-model}.
	For a given vertex $v \in [n]$, as $\theta_v$ increases, so does the probability edges lying on it appear in one or the other, but not both, of the current and next iterations of $G$.
	If an edge lying on $v$ is in $\vec{G}_t$ and it appears in $\vec{G}_t \triangle \vec{G}_{t+1}$, then that dyad will not appear in $\vec{G}_{t+1}$.
	Likewise, if a dyad lying on $v$ is not in $\vec{G}_t$ and it appears in $\vec{G}_t \triangle \vec{G}_{t+1}$, then that edge will appear in $\vec{G}_{t+1}$.
	Thus $G$ \emph{changes} the most in the neighborhoods of vertices $v$ with high values of $\theta_v$.
	We might say that $v$ is \introduce{loyal} if it has a low value of $\theta_v$ (near zero) and is \introduce{disloyal} if it has a high value of $\theta_v$.
\end{example}

\section{Concluding Remarks}\label{sec:conclusion}
	Our original goal was to develop goodness-of-fit tests for the \abbr{TERGM} examples in \textcite{hfx2010} using algebraic statistical techniques from \abbr{ERGM} analysis \autocite{Gross2022}.
	These techniques presuppose that the null model is an exponential family; for a stochastic process, what matters is the joint distribution of a finite sample.
	We needed \abbr{TERGM}s to have exponential-family representations with the same parameter as their transition probabilities because the transitions' parameters are of ultimate interest.
	However, \textcite{hfx2010} defined \abbr{TERGM}s as what we have been calling \emph{\abbr{CEF}s} (though not in this language).
	For concreteness, consider a \abbr{TERGM} $X = \{\vec{X}_t\}_{t\in\N}$, which is a Markov chain on $\Graphs[1]{n}$ whose transition probabilities have the \abbr{CEF} representation in \cref{eq:def-cef}.
	We had hoped that some straightforward calculations of the joint \abbr{PMF} using the Markov property would at least show that the statistic $\sum_{i=0}^{t-1} \vec{\tau}(\vec{X}_i, \vec{X}_{i+1})$ is sufficient for the transitions' parameter $\vec{\theta}$.

	The result of these calculations, which \textcite[Thm.~2.3.13, \pnfmt{33}]{Schwartz2021} presented, was that the normalizing term of $\Pr_{\vec{\theta}}(\vec{X}_1 = \vec{x}_1, \dotsc, \vec{X}_t = \vec{x}_t)$ depends on the observed data $\vec{x}_1, \dotsc, \vec{x}_t$.
	This does not happen in exponential-family representations.
	While not obvious from the formulae that \textcite{hfx2010} used to define their statistics $\vec{\tau}$, this dependence of the normalizing term on the observed data turned out to be because the log-partition function $\lnpart$ in \cref{eq:def-cef} depends on the previous state of the Markov chain.
	As \cref{thm:mef-exp-fam} turns out to say, to clear the dependence while still testing hypotheses on $\vec{\theta}$ requires that a \abbr{TERGM}'s transition probabilities have not just a \abbr{CEF} but an \emph{\abbr{MEF}} representation.

	The challenge remained of identifying when a given \abbr{TERGM} (or any \abbr{CEF}) is also an \abbr{MEF}.
	Permutation uniformity offers a nice solution because all p-uniform \abbr{CEF}s are \abbr{MEF}s (\cref{thm:uniform-cef-exp-fam-is-mef}), p-uniformity is discoverable directly from the sufficient statistics around which models are built (\cref{thm:puniform-suff-stat-implies-puniform-transition-matrix}), and, most of all, the \abbr{IID} sequences they provide (\cref{thm:uniform-chain}) are easy to interpret in the same terms as the original model (\cref{thm:uniform-cef-exp-fam}).
	In the case of a dyadically independent \abbr{TERGM}, it is particularly appealing to view a temporally dependent sequence of networks as a single observation of a multigraph (\cref{thm:suff-stat-dyadic-indep-multigraphs,thm:suff-stat-dyadic-indep-multigraph-joint-prob}).
	In this form, we expect algebraic statistical techniques will apply easily to develop goodness-of-fit tests for dyadically independent, p-uniform \abbr{TERGM}s and, perhaps with some work, to other \abbr{TERGM}s that are \abbr{MEF}s.

	Now that our main results are in hand, one can string them together to obtain a technique for analyzing \abbr{TERGM}s, namely a general recipe for dispensing with the joint likelihood's denominator's dependence on the observed data so that one may rewrite the likelihood in exponential-family form.
	Specifically, \cref{thm:mef-exp-fam} says that the joint distribution of an \abbr{MEF} has an exponential-family representation with the same parameter as the transition probabilities in \cref{eq:def-mef}.
	This permits the usual exponential-family \emph{computation} of the \abbr{MLE} for the joint likelihood's parameter, which is the same as the \emph{transitions'} parameter.
	\Cref{thm:uniform-chain} says that we can rewrite a Markov chain as a temporally independent sequence (irrespective of dyadic independence), permitting an intuitive \emph{interpretation} of the \abbr{MLE}.
	The \lcnamecref{thm:uniform-chain} associates a p-uniform \abbr{TERGM} with an \abbr{IID} sequence of \abbr{ERGM}s.
	We can take the multigraph union of $t$ samples from that sequence.
	This multigraph union is then a $t$-\abbr{ERMGM} with the same parameter as the original \abbr{TERGM} \autocite[26]{lehmann-casella-point-est}.
	However, if the distribution of the \abbr{IID} \abbr{ERGM}s is dyadically independent, then so is the $t$-\abbr{ERMGM} by \cref{thm:suff-stat-dyadic-indep-multigraphs,thm:suff-stat-dyadic-indep-multigraph-joint-prob}.

	Looking back at the literature, \textcite{hfx2010} proposed four statistics for transitions in the Markov chain, and we address each of them in \cref{sec:examples-lit}:
	Density is such that $\vec\tau$ already does not depend on the Markov chain's present state ($\vec{x}_i$ in the notation above).
	Stability does, but, because it is p-uniform, \cref{thm:mef-exp-fam} permits us to rewrite the joint likelihood of a stability-based \abbr{TERGM} in exponential-family form.
	Reciprocity and transitivity also do depend on $\vec{x}_i$, but our results prove that one \emph{cannot} rewrite the \abbr{TERGM}s based on them in exponential family form with the same parameter as the transitions.
	The new model in \cref{sec:loyalty-tergm} is p-uniform, thus \cref{thm:uniform-chain} applies, providing a more complex Markov-chain-transition sufficient statistic that leads to an exponential-family multigraph model with interpretable parameters.

\section*{Acknowledgements}

	William thanks Alessandro Rinaldo for his pivotal pointer to \textcite{Kuchler:1997vo}.
	The Air Force Office of Scientific Research's grant number \texttt{FA9550-14-1-0141} partially supported William's and Sonja's initial work on this project.
	The Simons Foundation's Collaboration Grant for Mathematicians number \texttt{854770} also partially supported Sonja's work.
	William, Sonja, and Hemanshu thank the three anonymous referees and associate editor for their reviews of an early draft of this report.

\printbibliography[heading=bibintoc]

\section*{Correspondence}
	\begin{quote}
		William K. Schwartz \texttt{<}\email{wkschwartz@gmail.com}\texttt{>} \\
		Secretariat Economists \\
		2121 K Street NW, Suite 1100 \\
		Washington, DC~20037
	\end{quote}

\clearpage
%TC:ignore
\appendix
	\appendixpage           % Show an "Appendices" header here...
	\noappendicestocpagenum % ...and, without showing a page number in the TOC,...
	\addappheadtotoc        % ...show "Appendices" in the TOC (and PDF sidebar?).

	\section{Two Equivalent Definitions of Permutation Uniformity}\label{sec:appendx-yanirosenblatt}
		\Textcite[\S~3]{rosenblatt-stationary-proc-as-shifts} first named and studied p-uniform chains, calling them simply \emph{uniform chains}.
		\Textcite[Def.~1.4]{markov-chains-as-random-walks} added the \emph{p-} in their definition of p-uniform chains, which they defined as Markov chains whose transition matrices $f$ satisfy \cref{eq:uniform-chain}.
		The following \lcnamecref{thm:uniform-chain-yano-definition} asserts the equivalence of \citeauthor{markov-chains-as-random-walks}'s definition and ours (derived from \citeauthor{rosenblatt-stationary-proc-as-shifts}'s).

		\begin{lemma}\label{thm:uniform-chain-yano-definition}
			$f$ is p-uniform under $\perm$ if and only if there exists a function $g \colon \St \to \mathscr{A}$ such that
			\begin{equation}\label{eq:uniform-chain}
				f(a, b) = g(\perm_a b) \qquad \text{for all } a,b \in \St.
			\end{equation}
			In this case, $g$ is unique up to permutation.
			If $f$ is a p-uniform stochastic matrix on $\St$, then $g$ is a stochastic vector, i.e., a \abbr{PMF}.
		\end{lemma}

		\begin{proof}
			\Cref{eq:uniform-chain} follows directly from the definition of permutation uniformity.

			Suppose $g, h \colon \St \to \mathscr{A}$ and $\{\perm_a\}_{a\in\St}$ and $\{s_a\}_{a\in\St}$ are sets of permutations such that $f(a, b) = g(\perm_a b) = h(s_a b)$ for all $a, b$.
			Then for any $b \in \St$, $g(b) = g(\perm_a \perm^{-1}_a b) = f(a, \perm^{-1}_a b) = h(s_a \perm^{-1}_a b)$.
			That is, $g$ is the composition of $h$ and the permutation $s_a \perm^{-1}_a$.

			The last part of the proof follows from the definition of a stochastic matrix.
		\end{proof}

	%%%%% Using \ref inside (sub)section titles
	%
	% Using amsmath and hyperref together, as we do, makes \cref and related
	% commands fail inside section titles. The recommended workaround is to use
	% \ref instead of \cref. This means manually entering the reference type
	% name. (The mode of failure seems to be no more severe than warnings and
	% malformed PDF bookmarks. Anyway, its usage is officially not supported.)
	%
	% cleveref 0.21.4, released 2018/03/27 and available at least as of MacTex 2021.
	% See the first bullet point under §14.2 on p. 26 of the documentation.
	\section{Markov Chains Induced by Random Functions and Theorem \ref{thm:uniform-chain}}\label{sec:appendx-induced}
		This \lcnamecref{sec:appendx-induced} translates \cref{thm:uniform-chain}'s language for p-uniform chains into \textcite{diaconis-freedman-1999}'s iterated random functions model that \cref{ex:modular-autoregression} alluded to.

		Let $\Omega$ be an arbitrary set.
		Fix some family $F = \{f_\omega\}_{\omega \in \Omega}$ of functions from $\St$ to itself indexed by $\Omega$.
		Let $\mu$ be a \abbr{PMF} defined on $\Omega$.
		Suppose $X$ is a Markov chain under some probability measure $\Pr$.
		$X$ is \introduce{induced by} $F$ and $\mu$ starting at $x_0 \in \St$ if $X_0 = x_0$ and, for $t \in \N$, $X_{t+1} = f_{W_{t+1}}(X_t)$, where $W = \{W_t\}_{t \in \N}$ is an \abbr{IID} sequence of $\Omega$-valued random variables with common law $\mu$ under $\Pr$.
		When $\Omega$ is discrete, we have that
		\[
			\Pr(X_{t+1} = b \given X_t = a)
			= \Pr(f_{W_{t+1}}(a) = b)
			= \sum_{\mathclap{\substack{\omega \in \Omega \\ f_\omega(a) = b}}} \mu(\omega).
		\]

		Adopting the notation of \cref{thm:uniform-chain}, let $X$ be a Markov chain p-uniform under $\perm$ with an \abbr{IID} sequence $Z$ with common law $\mu$.
		Setting $\Omega = \St$ and $f_a(b) = \perm^{-1}_b a$, we have that $X$ is induced by $F$ and $\mu$:
		\[
			X_{t+1}
			= \perm^{-1}_{X_t}Z_{t+1}
			= f_{Z_{t+1}}(X_t).
		\]
		In this case, $\{\omega \in \Omega \given f_\omega(a) = b\} = \{\perm_a b\}$, so that $\Pr(X_{t+1} = b \given X_t = a) = \mu(\perm_a b)$ as desired under p-uniformity.

		Permutation uniformity forces us to restrict the functions $f_a$.
		Since the $\perm_a$s are bijective, $a \mapsto f_a$ is injective in the set of functions $\St \to \St$, and $a \mapsto f_a(b)$ is bijective in $\St$.
		The latter assertion follows straight from the definition of $f_a$; to see the former, note that
		\[
			f_a = f_b
			\iff
			f_a(c) = f_b(c)\ \forall c
			\iff
			\perm^{-1}_c a = \perm^{-1}_c b\ \forall c
			\iff
			a = b.
		\]
		Other works that consider iterating random maps to form dependent sequences are \textcites[4]{Wu:2010jh}{markov-chains-as-random-walks}.

		\begin{ExampleContinued}{ex:modular-autoregression}
			Recall from the first part of the \lcnamecref{ex:modular-autoregression} \vpageref{ex:modular-autoregression} that $X_{t+1} = X_t + Z_{t+1} \pmod{n}$, where the $Z_t$s are uniform, \abbr{IID} random variables taking values zero or one, so that $X$ is a p-uniform chain under $\perm_i j = j - i \pmod{n}$ for each $i, j \in \St = \Z/n\Z$.

			In terms of iterating random functions, $f_j(i) = \perm^{-1}_i j = j + i \pmod{n}$, and we can write $X_{t+1} = f_{Z_{t+1}}(X_t)$.
			Notice that we apply only either $f_0(i) = i$ or $f_1(i) = i + 1 \pmod{n}$.
			$f_0$ and $f_1$ are Lipschitz continuous with Lipschitz constant one under the metric $\rho$ on $\St$ defined by $\rho(i, j) = \min\{j - i \pmod{n}, i - j \pmod{n}\}$, the shortest modular addition distance from $i$ to $j$ in either direction.
			\qed
		\end{ExampleContinued}

	\section{Dyadic Independence and Permutation Uniformity}
		Suppose $\vec{\tau} \colon \Graphs[t]{n} \times \Graphs[t]{n} \to \R^\ell$ is p-uniform under $\perm$, so that, by \cref{thm:uniform-chain-yano-definition}, there is a function $\vec{\nu} \colon \Graphs[t]{n} \to \R^\ell$ such that $\vec{\tau}(\vec{a}, \vec{b}) = \vec{\nu}(\perm_{\vec{a}} \vec{b})$ for all $\vec{a}, \vec{b} \in \Graphs[t]{n}$.
We say that this two-argument function $\vec{\tau}$ is \introduce{dyadditive} if $\vec{b} \mapsto \vec{\tau}(\vec{a}, \vec{b})$ is dyadditive for all $\vec{a} \in \Graphs[t]{n}$.
Can we extend dyadditivity from $\vec{\tau}$ to $\vec{\nu}$ or from $\vec{\nu}$ to $\vec{\tau}$?
Generally no.

\begin{example}
	For the case of simple graphs ($t = 1$) and scalar sufficient statistics ($\ell = 1$), we show that just because $\tau$ is dyadditive, it is not necessary that $\nu$ is.
	Suppose $n = 3$ and $\tau(\vec{a}, \vec{b}) = |E(\vec{b})|$, the number $\vec{1} \cdot \vec{b}$ of edges of $\vec{b}$, which is dyadditive (cf.~\cref{ex:dyadditive}).
	Set $\nu$ and $\perm$ so that
	\begin{align*}
		\nu(\vec{b}) &\coloneqq
		\begin{cases}
			|E(\vec{b})| & \text{if } \vec{b} \in \{\vec{0}, \vec{1}\} \\
			|E(\overline{\vec{b}})| & \text{else},
		\end{cases}
		&
		\perm_{\vec{a}}\vec{b} &\coloneqq
		\begin{cases}
			\vec{b}            & \text{if } \vec{b} \in \{\vec{0}, \vec{1}\} \\
			\overline{\vec{b}} & \text{else},
		\end{cases}
	\end{align*}
	for all $\vec{a}, \vec{b} \in \Graphs[1]{3}$.
	Hence $\tau(\vec{a}, \vec{b}) = \nu(\perm_{\vec{a}} \vec{b})$.

	Suppose by way of contradiction that $\nu$ is dyadditive.
	Per the comment after \cref{eq:dyadditive}, there are real numbers $q_1$, $q_2$, and $q_3$ such that $\nu(\vec{b}) = \sum_{f \in E(\vec{b})} q_f$.
	When $\vec{b}$ has one edge, $\nu(\vec{b}) = 2$, so $q_1 = q_2 = q_3 = 2$.
	But when $\vec{b}$ is the complete graph, we have $3 = \nu(\vec{b}) = q_1 + q_2 + q_3 = 2 + 2 + 2$, a contradiction.
	\qed
\end{example}

There is one case where we can guarantee that dyadditivity passes from $\vec{\tau}$ to $\vec{\nu}$: when one of the permutations is the identity permutation, such as in \cref{ex:puniform-symmetric-difference}.

\begin{proposition}\label{thm:uniform-suff-stat-factor-over-edges-ident}
	If $\vec{\tau}$ is dyadditive and, for some $\vec{x} \in \Graphs[t]{n}$, $\perm_{\vec{x}}$ is the identity permutation, then $\vec{\nu}$ is dyadditive.
\end{proposition}

\begin{proof}
	For some $\vec{\tau}_f$s and all $\vec{b}$s, $
		\vec{\nu}(\vec{b})
		= \vec{\tau}(\vec{x}, \perm^{-1}_{\vec{x}} \vec{b})
		= \vec{\tau}(\vec{x},                      \vec{b})
		= \sum_{f\in\Dyads{n}}\vec{\tau}_f(\vec{x},b(f))
	$.
\end{proof}

	\section{Proofs}\label{sec:appendx-proofs}
		This \lcnamecref{sec:appendx-proofs} contains proofs of some of the technical statements in the paper.
		The proofs of \cref{thm:factor-over-edges-dyadic-indep,thm:suff-stat-dyadic-indep-multigraphs} rely on the following \lcnamecref{thm:sum-product-exchange} to facilitate applying the distributive law.
		Induction on $|\mathcal{T}|$ provides a short proof whose tedious index chasing we spare the reader.

		\begin{lemma}\label{thm:sum-product-exchange}
			Let $\mathcal{T}$ be a non-empty, finite set; $\{B_i\}_{i \in \mathcal{T}}$ be a family of finite sets; $B \coloneqq \bigcup_{i \in \mathcal{T}} B_i$; $f_i \colon B_i \to \R$ for each $i \in \mathcal{T}$; and $\mathcal{F}_{\mathcal{T}} = \left\{y \in B^{\mathcal{T}} \given y_i \in B_i \text{ for all } i \in \mathcal{T}\right\}$.
			Then
			\[
				\prod_{i \in \mathcal{T}}\sum_{b \in B_i} f_i(b)
				= \sum_{y \in \mathcal{F}_{\mathcal{T}}}\prod_{i \in \mathcal{T}} f_i(y_i).
			\]
		\end{lemma}

		% See comment above on 'Using \ref inside (sub)section titles'. Here we
		% repeat that pattern enough that we create some new commands for it.
		{ % Limit scope of new commands
			\newcommand\appxproofheaderlevel\subsection
			\newcommand\proofsubsection[2]{\appxproofheaderlevel{Proof of #1 \ref{#2}}}

			\proofsubsection{Observation}{thm:puniform-suff-stat-implies-puniform-transition-matrix}\label{proof:puniform-suff-stat-implies-puniform-transition-matrix}
				\begin{proof}
					For all $a, b, c \in \St$ and all $\vec{\theta} \in \Theta$,
\begin{align}
	\label{eq:puniform-suff-stat-implies-puniform-transition-matrix-a}
	P_{\vec{\theta}}(a, \perm^{-1}_a b)
	&= \kappa(a, \perm^{-1}_a b)\exp\left(\vec{\eta}(\vec{\theta}) \cdot \vec{\tau}(a, \perm^{-1}_a b) - \lnpart(a, \vec{\theta})\right) \\
	\label{eq:puniform-suff-stat-implies-puniform-transition-matrix-c}
	&= \kappa(c, \perm^{-1}_c b)\exp\left(\vec{\eta}(\vec{\theta}) \cdot \vec{\tau}(c, \perm^{-1}_c b) - \lnpart(a, \vec{\theta})\right).
\end{align}
We set 1 equal to the sum of \cref{eq:puniform-suff-stat-implies-puniform-transition-matrix-a} over all $b \in \St$ and rearrange to obtain
\begin{equation}\label{eq:puniform-suff-stat-implies-puniform-transition-matrix-beta}
	\exp\left(\lnpart(a, \vec{\theta})\right)
	= \sum_{b \in \St} \kappa(a, \perm^{-1}_a b)\exp\left(\vec{\eta}(\vec{\theta}) \cdot \vec{\tau}(a, \perm^{-1}_a b) \right).
\end{equation}
Doing the same thing to \cref{eq:puniform-suff-stat-implies-puniform-transition-matrix-c} and replacing $a$ with $c$ in \cref{eq:puniform-suff-stat-implies-puniform-transition-matrix-beta} reveals that $\lnpart(a, \vec{\theta}) = \lnpart(c, \vec{\theta})$.
Thus \cref{eq:puniform-suff-stat-implies-puniform-transition-matrix-c} equals
\begin{align*}
	\kappa(c, \perm^{-1}_c b)\exp\left(\vec{\eta}(\vec{\theta}) \cdot \vec{\tau}(c, \perm^{-1}_c b) - \lnpart(c, \vec{\theta})\right) &
	= P_{\vec{\theta}}(c, \perm^{-1}_c b).
	\qedhere
\end{align*}

				\end{proof}

			\appxproofheaderlevel{Exponential-Family Assumptions}\label{proof:param-space}
				Here we prove the claim from \cpageref{itm:param-space-contains-open-set} after the proof of \cref{thm:puniform-mef-implies-puniform-suff-stat}.

				\newcommand{\ParamFuncAffineIndepEntries}[1]{%
	To prove the first implication, for $i \in [\ell]$, let $\vec{e}_i$ be the $i$th standard basis vector in $\R^\ell$.
	The open set contains an open ball centered at some vector $\vec{c}$ with some radius $r$.
	The list of $\ell + 1$ vectors $\vec{c}, \frac{r}{2}\vec{e}_1 + \vec{c}, \dotsc, \frac{r}{2}\vec{e}_\ell + \vec{c}$ is affinely independent.

	For the second implication, consider any $\ell + 1$ affinely independent vectors in $\vec{\eta}(\Theta)$, say, $\vec{\eta}(\vec{\theta}_0), \dotsc, \vec{\eta}(\vec{\theta}_\ell)$.
	Then $\vec{\eta}(\vec{\theta}_1) - \vec{\eta}(\vec{\theta}_0), \dotsc, \vec{\eta}(\vec{\theta}_\ell) - \vec{\eta}(\vec{\theta}_0)$ are linearly independent#1.
	Form an $\ell \times \ell$ matrix $A$ with these vectors as columns.
	Suppose $\vec{\delta} \in \R^\ell$ and $h \in \R$ are such that $\vec{\delta} \cdot \vec{\eta}(\vec{\theta}) = h$ for all $\vec{\theta} \in \Theta$.
	Then $\vec{\delta} \cdot \left(\vec{\eta}(\vec{\theta}_i) - \vec{\eta}(\vec{\theta}_0)\right) = 0$ for all $i \in [\ell]$, and hence $A^{\transp} \vec{\delta} = \vec{0}$.
	By the linear independence of $A$'s columns, $\vec{\delta} = \vec{0}$, and thus $h = 0$.
	Hence $\vec{\eta}$ has affinely independent entries.
}

				\ParamFuncAffineIndepEntries{}

			\proofsubsection{Lemma}{thm:factor-over-edges-dyadic-indep}\label{proof:factor-over-edges-dyadic-indep}
				\begin{proof}
					\newcommand{\thmFactorOverEdgesDyadicIndep}[1]{
	That $\vec{G}$ is a $\Graphs[t]{n}$-valued random variable if and only if $G(f)$ is a $\{0, \dotsc, t\}$-valued random variable follows directly from the definition of $\Graphs[t]{n}$.
	For some $\vec{\theta} \in \Theta$, define $\mu_{\vec{\theta}}$ as in \cref{eq:exp-fam} and $\mu^f_{\vec{\theta}}$ as in \cref{eq:ermgm-dyad}.
	The backward implication will follow if we can show that, for all $\vec{g} \in \Graphs[t]{n}$, $\prod_{f \in \Dyads{n}}\mu^f_{\vec{\theta}}(g(f)) = \mu_{\vec{\theta}}(\vec{g})$, where \cref{eq:dyadditive,eq:dyadically-multiplicative} define $\vec{\tau}$ and $\kappa$ in terms of $\{\vec{\tau}_f\}_{f \in \Dyads{n}}$ and $\{\kappa_f\}_{f \in \Dyads{n}}$, respectively.
	The forward implication will follow if we can show that, for all $f \in \Dyads{n}$,
	\[
		\sum_{\substack{\vec{g} \in \St \\ g(f) = m}} \mu_{\vec{\theta}}(\vec{g}) = \mu^f_{\vec{\theta}}(m),
	\]
	where \cref{eq:dyadditive,eq:dyadically-multiplicative} define $\{\vec{\tau}_f\}_{f \in \Dyads{n}}$ and $\{\kappa_f\}_{f \in \Dyads{n}}$ in terms of $\vec{\tau}$ and $\kappa$, respectively.
	Then the dyadic independence of $\vec{G}$ will follow from the equality established when proving the backward implication.

	#1

	\textbf{Forward Implication.}
	Fix an arbitrary dyad $f \in \Dyads{n}$.
	For tidiness we write $\St = \Graphs[t]{n}$.
	The probability that $G(f) = m \in \{0, \dotsc, t\}$, is
	\begin{equation*}
		\sum_{\substack{\vec{g} \in \St \\ g(f) = m}} \mu_{\vec{\theta}}(\vec{g})
		= \frac{
			\sum_{\substack{\vec{g} \in \St \\ \mathclap{g(f) = m}}} \kappa(\vec{g})\exp\left(\vec{\eta}(\vec{\theta}) \cdot \vec{\tau}(\vec{g}) \right)
		}{
			\sum_{r=0}^t \sum_{\substack{\vec{x} \in \St \\ \mathclap{x(f) = r}}} \kappa(\vec{x})\exp\left(\vec{\eta}(\vec{\theta}) \cdot \vec{\tau}(\vec{x}) \right)
		},
	\end{equation*}
	where we have used \cref{eq:exp-fam-log-partition}.
	Using \cref{eq:dyadditive,eq:dyadically-multiplicative}, define
	\begin{align*}
		u(\vec{g})
		\coloneqq \prod_{\substack{h \in \Dyads{n} \\ h \ne f}} \kappa_{h}(g(h))\exp\left(\vec{\eta}(\vec{\theta}) \cdot \vec{\tau}_{h}(g(h)\right).
	\end{align*}
	$\vec{\tau}$ is dyadditive and $\kappa$ is dyadically multiplicative, so
	{
		\newcommand{\param}{\vec{\eta}(\vec{\theta})}
		\newcommand{\expr}[1]{\kappa_{##1}(g(##1))\exp\left(\param \cdot \vec{\tau}_{##1}(g(##1))\right)}
		\begin{align*}
			\kappa(\vec{g})\exp\left(\param \cdot \vec{\tau}(\vec{g}) \right)
			&= \expr{f}u(\vec{g}).
		\end{align*}
	}
	The expression for $u(\vec{g})$ does not involve $g(f)$, so $u(\vec{g}) = u(\vec{x})$ regardless of whether $\vec{g}, \vec{x} \in \St$ have edge $f$ the same number of times.
	Consequently,
	\begin{equation}\label{eq:factor-over-edges-dyadic-indep-equal-sum}
		\sum_{\substack{\vec{g} \in \St \\ g(f) = m}} u(\vec{g})
		= \sum_{\substack{\vec{x} \in \St \\ x(f) = r}} u(\vec{x})
	\end{equation}
	for each $r \in \{0, \dotsc, t\}$.
	Factoring out this sum gives
	\begin{equation*}
		\frac{
			\sum_{\substack{\vec{g} \in \St \\ \mathclap{g(f) = m}}} \kappa(\vec{g})\exp\left(\vec{\eta}(\vec{\theta}) \cdot \vec{\tau}(\vec{g}) \right)
		}{
			\sum_{r=0}^t \sum_{\substack{\vec{x} \in \St \\ \mathclap{x(f) = r}}} \kappa(\vec{x})\exp\left(\vec{\eta}(\vec{\theta}) \cdot \vec{\tau}(\vec{x}) \right)
		}
		= \dfrac{
			\kappa_f(m)\exp\left(\vec{\eta}(\vec{\theta}) \cdot \vec{\tau}_f(m)\right) \sum_{\substack{\vec{g} \in \St \\ \mathclap{g(f) = m}}} u(\vec{g})
		}{
			\sum_{r=0}^t \kappa_f(r)\exp\left(\vec{\eta}(\vec{\theta}) \cdot \vec{\tau}_f(r)\right) \sum_{\substack{\vec{x} \in \St \\ \mathclap{x(f) = r}}} u(\vec{x})
		},
	\end{equation*}
	which, after canceling out \cref{eq:factor-over-edges-dyadic-indep-equal-sum} in the numerator and denominator, equals \cref{eq:ermgm-dyad}.

	\textbf{Backward Implication.}
	First off, for any $\vec{x} \in \Graphs[t]{n}$, we have
	\begin{align}
		\notag
		\prod_{f \in \Dyads{n}} \kappa_f(x(f)) \exp\left(\vec{\eta}(\vec{\theta}) \cdot \vec{\tau}_f(x(f))\right)
		&= \exp\left(\vec{\eta}(\vec{\theta}) \cdot \sum_{f \in \Dyads{n}}\vec{\tau}_f(x(f))\right) \prod_{f \in \Dyads{n}} \kappa_f(x(f)) \\
		\label{eq:factor-over-edges-dyadic-indep-prod-pass-through}
		&= \kappa(\vec{x}) \exp\left(\vec{\eta}(\vec{\theta}) \cdot \vec{\tau}(\vec{x})\right),
	\end{align}
	where we have defined $\vec{\tau}$ and $\kappa$ via \cref{eq:dyadditive,eq:dyadically-multiplicative}.

	Fix an arbitrary graph $\vec{g} \in \Graphs[t]{n}$.
	Then, using \cref{eq:factor-over-edges-dyadic-indep-prod-pass-through}, the (joint) probability that $\vec{G} = \vec{g}$ is
	\begin{align*}
		\prod_{h \in \Dyads{n}} \mu^h_{\vec{\theta}}(g(h))
		= \frac{
			\kappa(\vec{g}) \exp\left(\vec{\eta}(\vec{\theta}) \cdot \vec{\tau}(\vec{g})\right)
		}{
			\prod_{h \in \Dyads{n}} \sum_{r=0}^t \kappa_h(r) \exp\left(\vec{\eta}(\vec{\theta}) \cdot \vec{\tau}_h(r)\right)
		}.
	\end{align*}

	This matches \cref{eq:exp-fam} if we can show that the denominator equals $e^{\lnpart(\vec{\theta})}$.
	To that end, we exchange $\prod_{f \in \Dyads{n}}$ and $\sum_{r=0}^t$ using \cref{thm:sum-product-exchange}.
	In the language of \lcnamecref{thm:sum-product-exchange}, set $\mathcal{T} \coloneqq \Dyads{n}$, and $B_h \coloneqq \{0, \dotsc, t\}$ and $f_h(r) \coloneqq \exp\left(\vec{\eta}(\vec{\theta}) \cdot \vec{\tau}_h(r)\right)$ for each $r \in B_h$ and each $h \in \mathcal{T}$.
	Then $\mathcal{F}_{\mathcal{T}} = \Graphs[t]{n}$, and
	\begin{align}
		\label{eq:factor-over-edges-dyadic-indep-denom}
		\prod_{h \in \Dyads{n}} \sum_{r=0}^t \kappa_h(r) \exp\left(\vec{\eta}(\vec{\theta}) \cdot \vec{\tau}_h(r)\right)
		&= \sum_{\vec{x} \in \Graphs[t]{n}} \prod_{h \in \Dyads{n}} \kappa_h(x(h)) \exp\left(\vec{\eta}(\vec{\theta}) \cdot \vec{\tau}_h(x(h))\right) \\
		\notag
		&= \sum_{\vec{x} \in \Graphs[t]{n}} \kappa(\vec{x}) \exp\left(\vec{\eta}(\vec{\theta}) \cdot \vec{\tau}(\vec{x})\right)
		= \exp\lnpart(\vec{\theta}),
	\end{align}
	where the second equality follows from \cref{eq:factor-over-edges-dyadic-indep-prod-pass-through} and the last from \cref{eq:exp-fam-log-partition}.
}

					\thmFactorOverEdgesDyadicIndep{
						Both steps require the following fact.
						Setting the sum of \cref{eq:exp-fam} over $\St$ to one yields
						\begin{equation}\label{eq:exp-fam-log-partition}
							\lnpart(\vec{\theta})
							= \log\sum_{\vec{x} \in \St}\kappa(\vec{x})\exp\left(\vec{\eta}(\vec{\theta}) \cdot \vec{\tau}(\vec{x})\right).
						\end{equation}
					}
				\end{proof}

			\proofsubsection{Theorem}{thm:suff-stat-dyadic-indep-multigraphs}\label{proof:suff-stat-dyadic-indep-multigraphs}
				\begin{proof}
					First we show dyadic independence.
By the law of total probability,
\begin{equation*}
	\Pr(\vec{W} = \vec{w})
	= \sum_{\mathclap{\substack{x \in \Graphs[1]{n}^t \\ \sum_i \vec{x}_i = \vec{w}}}} \; \Pr(\vec{Z}_1 = \vec{x}_1, \dotsc, \vec{Z}_t = \vec{x}_t)
	= \sum_{\mathclap{\substack{x \in \Graphs[1]{n}^t \\ \sum_i \vec{x}_i = \vec{w}}}} \; \prod_{f \in \Dyads{n}} \prod_{i=1}^t \Pr(Z_i(f) = x_i(f))
\end{equation*}
since $Z$ is \abbr{IID} and dyadically independent.

To apply \vref{thm:sum-product-exchange} to swap the sum and product above, notice that $\mathcal{T} \coloneqq \Dyads{n}$ is a finite set.
For each $f \in \Dyads{n}$, take $B_f$ to be the set of indicator vectors for the time periods $\le t$ at which $f$ could enter the multigraph union: $B_f \coloneqq \{\vec{b} \in \{0, 1\}^t \given \sum_{i=1}^t b_i = w(f)\}$.
Further, take $h_f \colon B_f \to \R$ such that $h_f(\vec{b}) = \prod_{i=1}^t \Pr(Z_i(f) = b_i)$.
In the notation of \cref{thm:sum-product-exchange}, this makes $\mathcal{F}_{\Dyads{n}} = \{x \in \Graphs[1]{n}^t \given \sum_{i=1}^t \vec{x}_i = \vec{w}\}$, so we may interchange the sum and the product as follows.
\begin{align}
	\label{eq:suff-stat-dyadic-indep-multigraphs-intermediate}
	\Pr(\vec{W} = \vec{w})
	&= \prod_{f \in \Dyads{n}} \sum_{\substack{\vec{b} \in \{0,1\}^t \\ \sum_i b_i = w(f)}} \prod_{i=1}^t \Pr(Z_i(f) = b_i) \\
	\tag{Z \text{ is \abbr{IID}}}
	&= \prod_{f \in \Dyads{n}} \sum_{\substack{\vec{b} \in \{0,1\}^t \\ \sum_i b_i = w(f)}} \Pr(Z_1(f) = b_1, \dotsc, Z_t(f) = b_t) \\
	\tag{\text{Law of total prob.}}
	&= \prod_{f \in \Dyads{n}} \Pr\left(\sum_{i=1}^t Z_i(f) = w(f)\right)
	=  \prod_{f \in \Dyads{n}} \Pr(W(f) = w(f)).
\end{align}
Therefore the multiplicities of each dyad in $\vec{W}$ are independent of each other.

To prove \cref{eq:suff-stat-dyadic-indep-multigraphs}, let $\mu_f \coloneqq \Pr(Z_i(f) = 1)$ for each $f \in \Dyads{n}$.
Since $Z$ is \abbr{IID} and dyadically independent,
\[
	\Pr(Z = z)
	= \prod_{f\in \Dyads{n}} \prod_{i=1}^t \Pr(Z_i(f) = z_i(f))
	= \prod_{f\in \Dyads{n}} \mu_f^{w(f)}(1-\mu_f)^{t-w(f)}.
\]
Likewise, from \cref{eq:suff-stat-dyadic-indep-multigraphs-intermediate} and the combinatorial definition of the binomial coefficient, we have
\begin{align*}
	\Pr(\vec{W} = \vec{w})
	&= \prod_{f \in \Dyads{n}} \sum_{\substack{b \in \{0,1\}^t \\ \sum_i b_i = w(f)}} \prod_{i=1}^t \Pr(Z_i(f) = b_i)
	= \prod_{f \in \Dyads{n}}  \sum_{\substack{b \in \{0,1\}^t \\ \sum_i b_i = w(f)}} \mu_f^{w(f)}(1-\mu_f)^{t-w(f)} \\
	&= \prod_{f \in \Dyads{n}} \binom{t}{w(f)} \mu_f^{w(f)}(1-\mu_f)^{t-w(f)}
	= \Pr(Z = z)\prod_{f \in \Dyads{n}} \binom{t}{w(f)}.
	\qedhere
\end{align*}

				\end{proof}
		}
%TC:endignore
\end{document}